\pdfpageattr {/Group << /S /Transparency /I true /CS /DeviceRGB>>}
\documentclass{article}
\usepackage[margin=1in]{geometry}
\usepackage{amsmath}
\usepackage{amssymb}
\usepackage{xfrac}
\usepackage{authblk}
\usepackage{longtable}
\usepackage{xfrac}
\usepackage{tkz-graph}
\usepackage{tkz-graph}
\usepackage{array}
\usepackage{rotating}
\newcolumntype{z}{>{$\vcenter\bgroup\hbox\bgroup}c<{\egroup\egroup$}}
\PassOptionsToPackage{cmyk}{xcolor}

\DeclareMathOperator{\Tr}{Tr} 

\title{Superelliptical laws for complex networks}

\author[1,2]{Stefano Allesina\thanks{sallesina@uchicago.edu}}
\author[1]{Elizabeth Sander} 
\author[1]{Matthew J. Smith}
\author[1]{Si Tang}
\affil[1]{Department of Ecology \& Evolution, University of Chicago}
\affil[2]{Computation Institute, University of Chicago}

\begin{document}

\maketitle

\begin{abstract}
  All dynamical systems of biological interest---be they food webs,
  regulation of genes, or contacts between healthy and infectious
  individuals---have complex network structure. Wigner's semicircular
  law and Girko's circular law describe the eigenvalues of systems
  whose structure is a fully connected network. However, these laws
  fail for systems with complex network structure. Here we show that
  in these cases the eigenvalues are described by superellipses. We
  also develop a new method to analytically estimate the dominant
  eigenvalue of complex networks.
\end{abstract}

Eigenvalues and eigenvectors are central to science and
engineering. They are used to assess the stability of dynamical
systems\cite{allesina2012stability}, the synchronization of
networks\cite{barahona2002synchronization}, the occurrence of
epidemics of infectious
diseases\cite{pastor2001epidemic,chakrabarti2008epidemic,youssef2011individual,li2012susceptible,diekmann2010construction},
and the ranking of web-pages\cite{bryan200625}.

Two connected results describe the eigenvalue distribution of large
matrices with random coefficients: Wigner's semicircle
law\cite{wigner1958distribution} (for symmetric matrices), and Girko's
circular law\cite{girko1985circular} (for asymmetric ones).

These laws have been applied in a wide range of disciplines, from
number theory\cite{terras2011zeta} to
ecology\cite{allesina2012stability}. However, they fall short of
describing the eigenvalues of matrices with complex network
structure\cite{farkas2001spectra}. Such matrices are ubiquitous in the
analysis of man-made and biological
systems\cite{barabasi1999emergence,newman2003structure}.

Here we analyze matrices whose underlying structure is a sparse random
network with arbitrary degree distribution.  We show that the
eigenvalue distributions of these matrices are described by
superellipses. We solve the case of symmetric matrices by generalizing
Wigner's law, and extend Girko's law to matrices with regular-graph
structure (where all nodes have the same degree). We also propose a
new method to approximate the dominant eigenvalue of a complex
network, improving upon current
results\cite{chung2003spectra,nadakuditi2013spectra}.

Wigner's semicircle law\cite{wigner1958distribution} states that the
density of the eigenvalues of a large, symmetric random matrix whose
entries are sampled from a normal distribution $\mathcal
N(0,\sigma^2)$ is described by a semicircle. Similarly, Girko's
circular
law\cite{ginibre1965statistical,girko1985circular,tao2010random}
states that the eigenvalues of asymmetric matrices with
normally-distributed entries are approximately uniformly distributed
in a circle. If there is a nonzero pairwise correlation between the
off-diagonal elements of the matrix, the eigenvalues are approximately
uniformly distributed in an
ellipse\cite{girko1985,sommers1988spectrum,Naumov20xx}. These laws
have found applications in a wide array of disciplines, including
quantum physics\cite{guhr1998random},
ecology\cite{allesina2012stability} and number
theory\cite{terras2011zeta}, and were recently proved to be
universal\cite{tao2010random,Naumov20xx} -- i.e., they hold under very
mild conditions on the distribution of the matrix entries.

Wigner's and Girko's laws hold for matrices whose underlying network
structure is a completely connected graph: as $s$, the size of the
graph, goes to infinity, the distribution of the eigenvalues converges
to the corresponding law. They also hold for not-completely-connected
graphs\cite{tao2010random}, as long as the number of connections per
node $s p$ (where $p$ is the proportion of nonzero entries in the
matrix) goes to infinity as $s \to \infty$.

However, the laws do not describe the eigenvalues of matrices with
complex network structure\cite{farkas2001spectra}. These networks are
typically sparse with a very heterogeneous degree
distribution\cite{newman2003structure}; moreover, for most complex
networks, we expect the average number of connections per node to
approach a constant as the network grows\cite{farkas2001spectra}. In
this case, as $s \to \infty$, $s p \to k$. The cap on the average
number of connections per node can arise from spatial or temporal
constraints. For example, in an ecological system organisms need to
spatiotemporally co-occurr in order to interact. The cap is also
evident in empirical data: Facebook counted around 56 million active
users in 2008\cite{backstrom2012four}, each with an average degree
(number of friends) of about 76, and while the number of users grew to
more 562 million in 2011, the average degree grew only to 169.

Our goal is to extend the circular laws above to the case of large
matrices with complex network structure, whose average degree is $k
\ll s$. We show that in such cases the eigenvalue distributions are
described by superellipses ($\sfrac{|x|^n}{a^n}
+ \sfrac{|y|^n}{b^n} \leq 1$, where $x$ is the real part of an
eigenvalue and $y$ is its imaginary part; for $n = 2$, we recover the
equation for an ellipse).

For symmetric matrices with normally-distributed entries (the analog
of Wigner's case), the density of the eigenvalues is described by a
semi-superellipse -- for any degree distribution of the underlying
network. For asymmetric matrices (Girko's case) whose structure is a
random $k$-regular graph (i.e., all nodes have the same degree), we
find that the eigenvalues are approximately uniformly distributed in a
superellipse. For other network structures, the distribution is still
described by superellipses, but is not uniform.

\section*{Results}
\subsection*{Symmetric Matrices}

\begin{sidewaysfigure}
  \includegraphics[width
    = \linewidth]{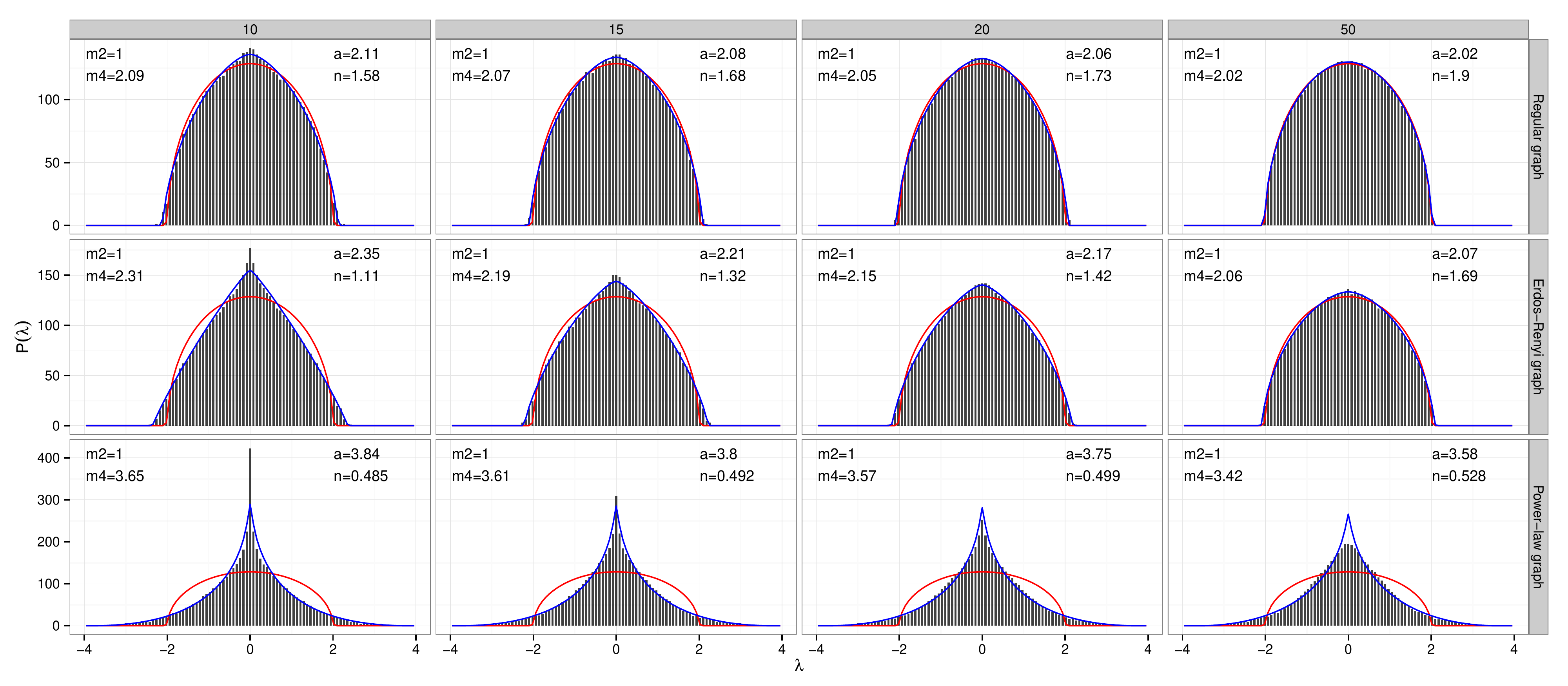} \caption{\textit{Probability
    density function for the eigenvalues of symmetric matrices whose
    underlying structure is a complex network. The network has average
    degree $k$ (columns), and degree distribution determined by
    different algorithms (rows). For a given degree distribution, the
    network is then built using the configuration
    model\cite{molloy1995critical,newman2003structure}. The values of
    the nonzero elements are sampled from a symmetric bivariate normal
    distribution centered at zero (SI). The red line marks the
    prediction of Wigner's semicircle
    distribution\cite{wigner1958distribution}. The blue line shows the
    semi-superelliptical distribution when $\mu_2$ and $\mu_4$ (top
    left in the panel) are used to solve the equations determining its
    parameters ($a$ and $n$ in the panels). Each graph is obtained by
    computing the eigenvalues of a single matrix of size $5000 \times
    5000$.}}
\end{sidewaysfigure}

We analyze $s \times s$ matrices with $0$ on the diagonal, and
off-diagonal pairs $(M_{ij},M_{ji})$ obtained by multiplying the
corresponding entries of two matrices $(M_{ij},M_{ji}) = (A_{ij},
A_{ji}) \cdot (N_{ij}, N_{ji})$. $A$ is the adjacency matrix of a
random undirected graph---with a given degree distribution---built
using the configuration
model\cite{molloy1995critical,newman2003structure}. The use of the
configuration model is important, as it ensures that the networks do
not typically have unwanted ``secondary structures'' (e.g., modules,
bipartite or lattice structure) that would affect results. $N$ is a
matrix whose off-diagonal pairs $(N_{ij}, N_{ji})$ are sampled from a
bivariate normal distribution $(X, Y)$, with $\mathbb E[X] = \mathbb
E[Y] = 0$, $\mathbb E[X^2] = \mathbb E[Y^2] = \sfrac{1}{k}$, and
$\mathbb E[X Y] = \sfrac{\rho}{k}$, where $k$ is the average degree of
the network and $-1 \leq \rho \leq 1$ is Pearson's correlation
coefficient. The choice of parameters ensures that for $k \to \infty$,
we recover the type of matrices studied by Wigner and Girko. Our
results also hold for non-normal distributions (e.g., uniform, SI).

We start with the analog of Wigner's case, in which matrices are
symmetric ($\rho = 1$). In this case, all eigenvalues are real, and
for $k \to \infty$ we recover Wigner's semicircle probability
distribution function:

\begin{equation}
  \Pr(\lambda = x) = P(x) = \frac{2 \sqrt{(2r)^2 - x^2}}{\pi (2r)^2}
\end{equation}
The variance of this distribution is a function of $r$: $\mu_2(r) =
\int x^2 P(x) dx = r^2$. Because the variance of the eigenvalues of a matrix
with diagonal zero is $\Tr(M^2)/s = \rho = \mu_2(r)$, given that in
our matrices $\rho =1$, then $r = 1$ and thus the horizontal radius is
$a = 2r = 2$. We next generalize Wigner's formula to the case where
$k \ll s$, which leads to a semi-superelliptical distribution:

\begin{equation}
  P(x) = \frac{2 \sqrt[n]{(2r)^n - x^n}}{4 (2r)^2 \Gamma(1 +
    \sfrac{1}{n})^2 \Gamma(1 + \sfrac{2}{n})^{-1}}
\end{equation}
\noindent where the numerator is the superelliptical equivalent of
Wigner's formulation, and the denominator is the area of a
superellipse with $a = b = 2r$. In this case, we need to solve for two
parameters, $r$ and $n$. Hence, we write equations for the second
($\mu_2(r,n) = \int x^2 P(x) dx = \Tr(M^2)/s$) and fourth ($\mu_4(r,n)
= \int x^4 P(x) dx = \Tr(M^4)/s$) central moments of the eigenvalue
distribution ($\mu_3(r, n) = 0$, due to symmetry), thereby obtaining
the values of $n$ and $r$ (SI).

In Figure 1, we show numerical simulations in which we take a single
$5000 \times 5000$ matrix, whose network structure is determined by
the average degree $k$ (columns) and a specific algorithm used to
construct the degree distribution (rows). In all cases, the density of
the eigenvalues is described by a semi-superellipse, which captures
the tails especially well. This is important, given the role of
dominant eigenvalues in determining the properties of dynamical
systems. The distribution tends to underestimate (small $k$) or
overestimate (large $k$) the number of zeros, especially for very
skewed degree distributions -- an effect similar to that found for
small matrices in Girko's circular law\cite{tao2010random}.

\subsection*{Asymmetric Matrices}

\begin{sidewaysfigure}
  \includegraphics[width =
  0.9\linewidth]{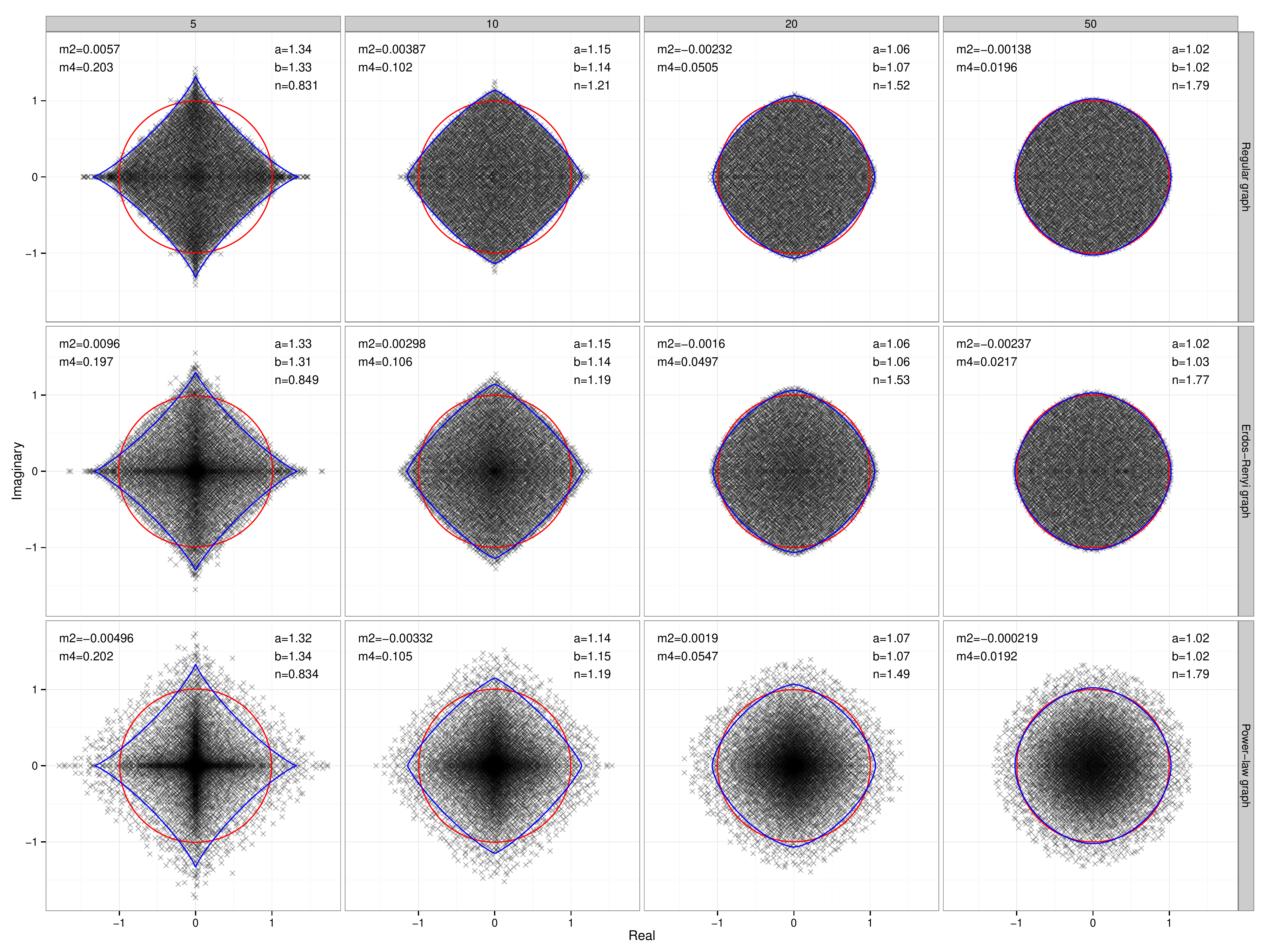} \caption{\textit{Distribution
  of the eigenvalues of asymmetric matrices ($\mathbb E[\rho] = 0$) in
  the complex plane. As in Figure 1, the columns specify the average
  degree of the network, and the rows specify the algorithm used to
  build the degree distribution. The red line is obtained using
  Girko's circular
  law\cite{ginibre1965statistical,girko1985circular,tao2010random}.
  The blue line is obtained by solving for $a$, $b$ and $n$ (top right
  in the panels) for the superellipse using the second and fourth
  moments of the eigenvalue distribution (top left in the panels). The
  superelliptical distribution is expected to accurately describe the
  case with $k$-regular graph structure, for which the eigenvalues are
  approximately uniform in the superellipse. Each graph is obtained
  computing the eigenvalues of a single matrix of size $5000 \times
  5000$.}}
\end{sidewaysfigure}

\begin{sidewaysfigure}
\includegraphics[width =
  0.9\linewidth]{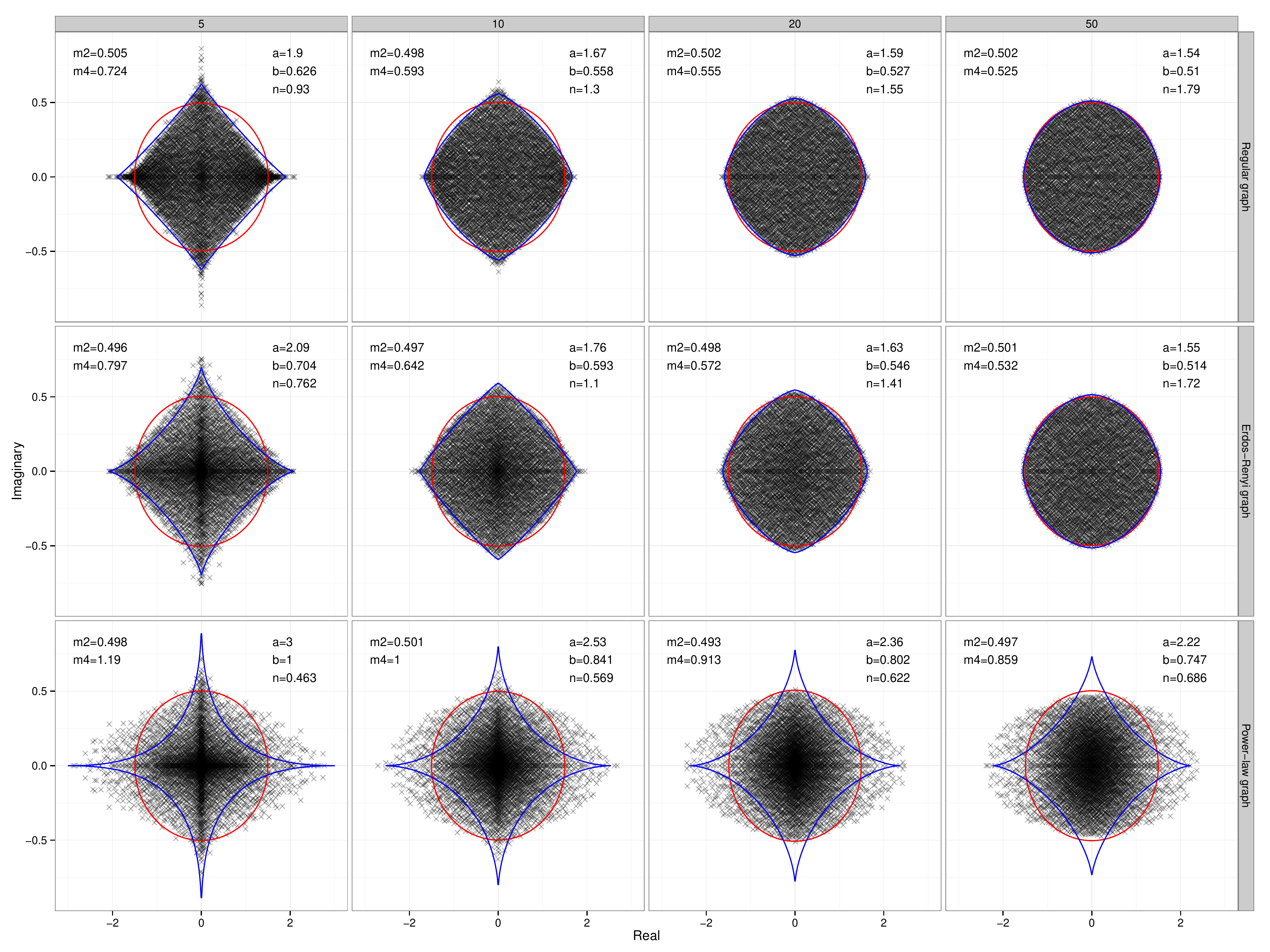}
\caption{\textit{As Figure 2, but with positive correlation $\mathbb
    E[\rho] = 0.5$. In this case the red line describes the elliptical
    law\cite{girko1985,sommers1988spectrum,Naumov20xx}.}}
\end{sidewaysfigure}

We now move to the case in which $\rho \neq 1$ (Figures 2 and 3). For
$k \to \infty$ and $\rho = 0$, we recover Girko's circular law, which
states that the eigenvalues are uniformly distributed in the unit
circle. For $k \to \infty$ and $-1< \rho <1$, the eigenvalues are
uniformly distributed in an ellipse with horizontal radius $a = 1 +
\rho$ and vertical radius $b = 1 -\rho$.

Although all the eigenvalue distributions appear to be described by
superellipses in the complex plane, only matrices with $k$-regular
graph structure have a distribution that is close to uniform. For
these matrices, we can approximate the p.d.f. as:

\begin{equation}
  \Pr(\Re(\lambda) = x, \Im(\lambda) =y ) = P(x,y) = \frac{1}{4 a b
    \Gamma(1 + \sfrac{1}{n})^2 \Gamma(1 + \sfrac{2}{n})^{-1}}
\end{equation}
Setting $a = r (1 + \rho)$ and $b = r (1-\rho)$ (SI), we can replicate
the approach above and write the second and fourth central moments as
$\mu_2(r,n) = \iint (x^2 - y^2) P(x,y) dy dx = \Tr(M^2)/s = \rho$ and
$\mu_4(r, n) = \iint (x^4 + y^4 - 6x^2y^2) P(x,y) dy dx = \Tr(M^4)/s$,
respectively. Solving these equations, we obtain $n$ and $r$ (SI).

\subsection*{Adjacency Matrices}
Many applications deal with matrices that do not comply with the
strict requirements we set above. For example, adjacency matrices of
undirected graphs have entries whose value is either zero or one. As
such, the mean of the matrix is not zero, and thus the eigenvalues do
not follow the semi-superellipse introduced above.

Consider a graph generated by the configuration model with arbitrary
degree distribution: the density of all eigenvalues but the dominant
one follows a semi-superellipse, while the dominant eigenvalue lies to
the right of the semi-superellipse (SI). We can exploit this fact to
accurately estimate the value of the dominant eigenvalue (Figure
4). 

One notable characteristic of semi-superelliptical distributions is
that they are symmetric about the mean. As such, the distribution
obtained by taking all the eigenvalues but the dominant one should
have odd central moments $\tilde{\mu}_{2z + 1} \approx 0$. Hence, one
can solve a system of equations of the form $\Tr(A^z)
= \sum_j \lambda_j^z = (s - 1)\tilde{\mu'}_{z} + \lambda^z_1$ (where
$\tilde{\mu'}_{z}$ is the $z^{th}$ raw moment of the distribution
obtained removing the dominant eigenvalue) by assuming, for example,
that $\tilde{\mu}_{3} =0$ or that $\tilde{\mu}_{5} =0$ (SI).

We contrast the two approximations obtained by setting the $3^{rd}$ or
$5^{th}$ central moments to zero and the fantastically simple one
proposed by Chung \textit{et al.}\cite{chung2003spectra}, $\lambda_1
\approx \overline{k^2} / \overline{k}$ (i.e., the average of the
squared degrees divided by the average degree).  Chung's approximation
holds as long as the minimum degree in the network is not too small
compared to the mean degree, and was independently obtained by
Nadakuditi \& Newman\cite{nadakuditi2013spectra} using free
probability. Figure 4 shows that the approximation based on the fifth
moment works better than the others for non-regular graphs.

\begin{figure}
  \begin{centering}
  \includegraphics[width = 0.75\linewidth]{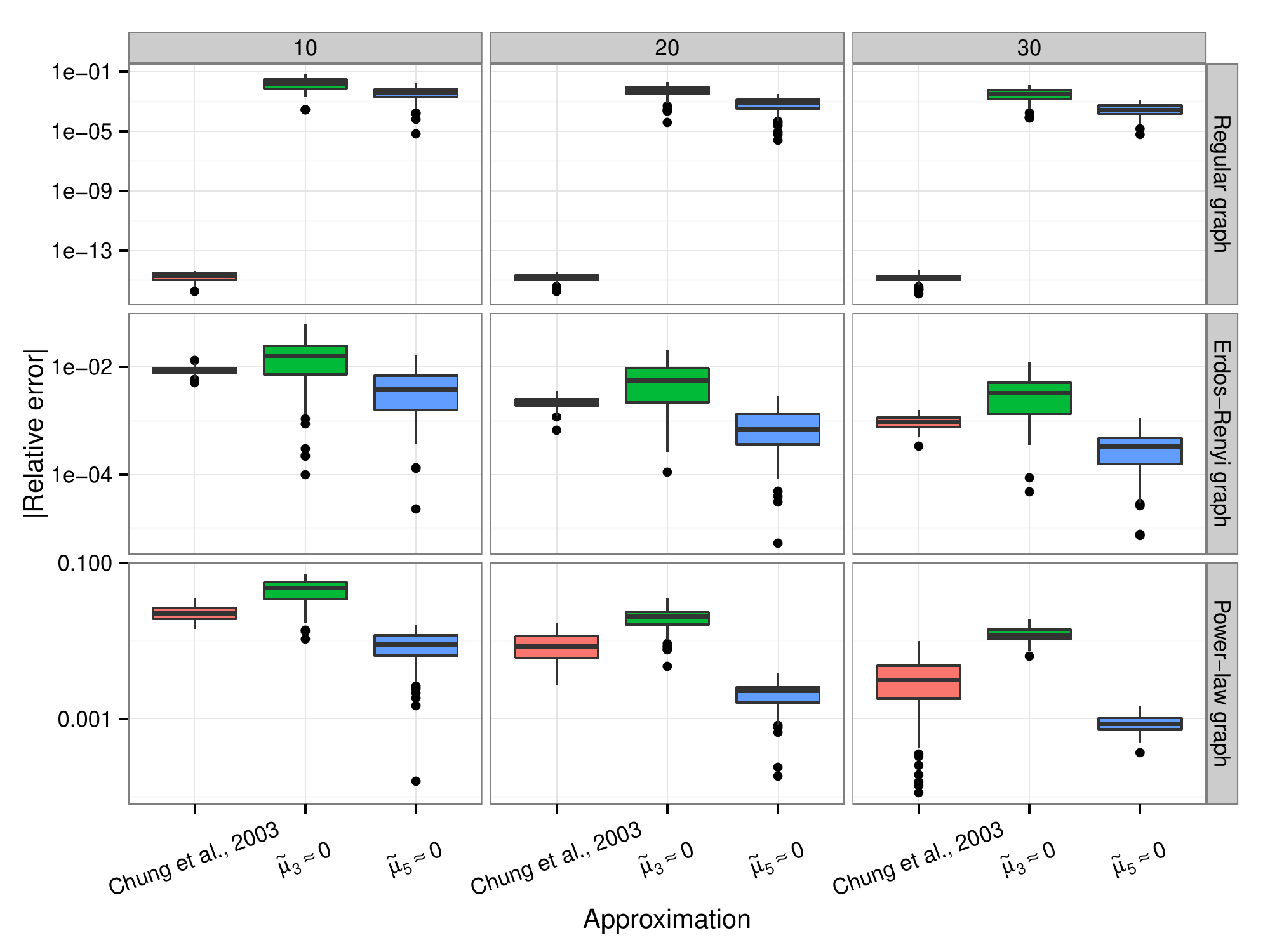}
  \caption{\textit{Approximation of the dominant eigenvalues of
      matrices built using the configuration model and choosing the
      degree distribution according to the algorithm specified in the
      rows. The columns stand for the desired average degree. The
      x-axis specifies the approximation used to obtain
      $\lambda_1^\text{approx}$, while the y-axis represents the
      absolute value of the relative error $|(\lambda_1^\text{approx}
      - \lambda_1)|/\lambda_1$. }}
      \end{centering}
\end{figure}

\section*{Discussion}

A semi-superellipse approximates the density of the eigenvalues of the
symmetric matrix $M$. When $M$ is asymmetric, the eigenvalues fall in
superellipses in the complex plane. When the structure of $M$ is a
random $k$-regular graph, then the distribution is approximately
uniform and we can estimate the parameters of the superellipse.

The spectrum of the adjacency matrix of a random graph with arbitrary
degrees can be described by a ``semi-superellipse plus $\lambda_1$''
distribution. Because a semi-superellipse is symmetric about the mean,
we can analytically approximate the value of $\lambda_1$. This allows,
for example, a more accurate prediction of the occurrence of epidemics
in simple Susceptible-Infected-Susceptible and
Susceptible-Infected-Recovered models that incorporate an explicit
network of contacts between
individuals\cite{chakrabarti2008epidemic,pastor2001epidemic,youssef2011individual,li2012susceptible},
for which the epidemic threshold is defined by $1/\lambda_1$.

Interestingly, we can connect the value of $\lambda_1$ with the
presence of small motifs\cite{milo2002network} in the network. Our
approximation of the dominant eigenvalue of adjacency matrices
produced by the configuration model shows that the dominant eigenvalue
of a network of a given size and connectivity strongly depends on the
number of triangles (approximation using $\tilde{\mu}_{3}$) and
pentagons (using $\tilde{\mu}_{5}$) it contains.

We also considered departures from the configuration model (SI). We
chose the configuration model in order to prevent the emergence of
``secondary structures'' in the network, but the ultimate test for
these methods is to approximate the behavior of real, empirical
networks, which are known to contain interesting structural
features\cite{newman2003structure}.

For symmetric matrices parametrized using the structure of networks of
biological interest, the density of the eigenvalues is captured by the
semi-superelliptical distribution. As expected, the asymmetric case is
not well-described by the superelliptical distribution derived for
$k$-regular graphs. Finally, in this case, the approximation of the
dominant eigenvalue continues to perform very well, and better than
current methods.

The results presented here open the door for the analytic study of
large dynamical systems with complex network structure.

\clearpage
\part*{Supporting Information}
\renewcommand{\thetable}{S\arabic{table}}   
\renewcommand{\thefigure}{S\arabic{figure}}
\renewcommand{\theequation}{S\arabic{equation}}
\setcounter{figure}{0}
\setcounter{equation}{0}
\section*{List of symbols}

For reference, we provide a list of the main symbols used in the
following sections.

\begin{longtable}{p{0.15\linewidth}p{0.85\linewidth}}
{Symbol} & {Description} \\ \hline
$\mathcal N(\boldsymbol \mu, \boldsymbol \Sigma)$ & Bivariate normal
distribution with mean $\boldsymbol \mu$ and covariance matrix
$\boldsymbol \Sigma$.\\
$\mathbb E[X]$ & Expectation of the random variable $X$.\\
$\text{Var}[X]$ & Variance of $X$.\\
$\rho$ & Pearson's correlation coefficient.\\
$\mathcal U(-t,t)$ & Uniform distribution on the interval $[-t, t]$. \\

$\mu'_z$ & $z^{th}$ raw moment (moment about the origin) of a
distribution. We sometimes write $\mu'_z(r,n)$ to stress that the
moment depends on the parameters of the distribution.\\
$\mu_z$ & $z^{th}$ central moment (moment about the mean) of a
distribution.\\

$A$ & The adjacency matrix of an undirected graph.\\
$A_{ij}$ & Coefficient of the adjacency matrix.\\
$N$ & A matrix whose entries $(N_{ij}, N_{ji})$ are sampled from a bivariate normal distribution.\\
$U$ & A matrix whose entries are sampled from a uniform distribution.\\
$M$ & Matrix obtained by multiplying the elements of $A$ and $N$ (or
$U$).\\
$\Tr(M)$ & Trace of the matrix $M$.\\
$M^z$ & Matrix $M$ raised to the $z^{th}$ power.\\

$k$ & The average degree of the nodes in a network described by $A$.\\
$s$ & Size of the network.\\

$\lambda_i$ & $i^{th}$ eigenvalue of a matrix.\\
$i$ & $\sqrt{-1}$.\\

$a$ & Horizontal radius of a superellipse.\\
$b$ & Vertical radius of a superellipse.\\
$r$ & $(a + b)/2$.\\
$n$ & Shape parameter of a superellipse.\\

$\Gamma(\cdot)$ & Gamma function.

\end{longtable}

\section*{Superelliptical distributions}
\subsection*{Matrices and Graphs.}

We analyze matrices $M$ of size $s \times s$ that are obtained by
multiplying the elements of two matrices: $M_{ij} = A_{ij}
N_{ij}$. $A$ is the adjacency matrix of an undirected, simple graph
without self-loops, generated by the configuration
model\cite{molloy1995critical,newman2003structure} (for a particular
degree distribution). The nodes in the graph have average degree
$k$. We consider the case of integer $k$, with $k>1$, so that the
graph is almost surely connected, and $k \leq (s - 1)$, as the graph
does not contain self-loops. Thus, the adjacency matrix contains
exactly $s k$ nonzero coefficients. $N$ is a matrix whose elements
$(N_{ij}, N_{ji})$ are sampled from a normal bivariate distribution
$\mathcal N(\boldsymbol \mu, \boldsymbol \Sigma)$, with

\begin{equation}
  \boldsymbol \mu = \left[
    \begin{array}{c}
    0 \\
    0
    \end{array}
    \right]
\end{equation}
\noindent and

\begin{equation}
  \boldsymbol \Sigma = \left[
    \begin{array}{cc}
    \frac{1}{k} & \frac{\rho}{k} \\
    \frac{\rho}{k} & \frac{1}{k}
    \end{array}
    \right]
\end{equation}
Clearly, $\mathbb E [M_{ij}] = 0$, and $\text{Var}[M_{ij}] = \mathbb E
[M_{ij}^2] = \sfrac{1}{s}$. For $k \to \infty$, each value of the
Pearson's correlation coefficient $\rho$ determines which of the
following well-known laws describes the distribution of the
eigenvalues of $M$:

\begin{table}[h]
  \begin{center}
  \begin{tabular}{c|l}
    $\rho$ & Law\\
    \hline
    $1$ & Wigner's semicircle law\cite{wigner1958distribution}\\
    $0$ & Girko's circular law\cite{ginibre1965statistical,girko1985circular,tao2010random}\\
    $-1 < \rho < 1$ & Elliptical law\cite{girko1985,sommers1988spectrum,Naumov20xx}
  \end{tabular}
  \end{center}
\end{table}
We want to study the case in which $k \ll s$, and the structure of the
graph $A$ is generated using different degree distributions. In
particular, we analyze degree distributions obtained from $k$-regular
random graphs, Erd\H{o}s-R\'enyi random graphs\cite{erdHos1959random},
and Power-law (scale-free) graphs\cite{barabasi1999emergence}. All
graphs were generated using the \texttt{igraph} library\cite{igraph}
for the statistical software \texttt{R}\cite{R}. For Power-law graphs,
undirected networks were obtained using the routine
\texttt{static.power.law.game} with a power-law exponent of $2.5$.

To obtain higher precision, we transform the nonzero coefficients of
the matrix $M$ to ensure that $\overline{M}_{ij} = 0$ and
$\overline{M^2}_{ij} = \sfrac{1}{s}$.

\subsection*{Superellipses.}
\begin{figure}
  \begin{centering}
  \includegraphics[width
  = 0.7\linewidth]{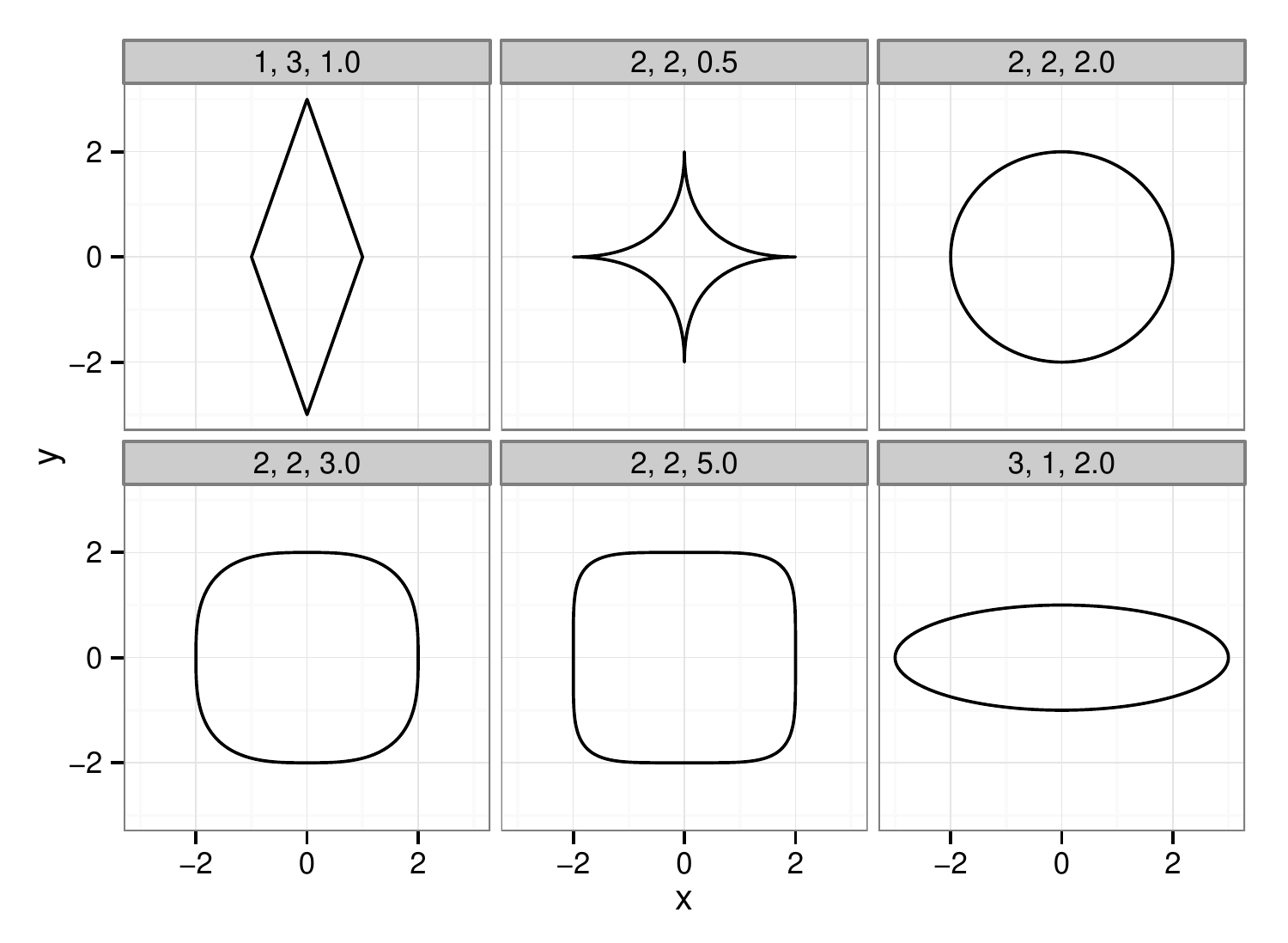} \caption{\textit{Shapes of
  $\sfrac{|x|^n}{a^n} + \sfrac{|y|^n}{b^n} = 1$, for different values
  of $a$, $b$ and $n$ (panel titles). }} \label{Super}
  \end{centering}
\end{figure}

As we will show, the eigenvalues of matrices with complex network
structure are described by superellipses. A superellipse is defined by
the equation:

\begin{equation}
  \frac{|x|^n}{a^n} + \frac{|y|^n}{b^n} = 1
\end{equation}
For $n = 2$ we recover the equation describing an ellipse (a circle
when $a = b$). The area of a superellipse is $4 a b \Gamma(1
+\sfrac{1}{n})^2 \Gamma(1 + \sfrac{2}{n})^{-1}$. Superellipses can
take dramatically different shapes depending on the parameter values
(Figure~\ref{Super}).

\subsection*{Moments of the eigenvalue distribution.}

A superellipse is defined by three parameters: $a$, $b$ and $n$.  When
deriving the parameters, we will use some well-known identities
relating the traces of the powers of a matrix to its eigenvalue
distribution.

In particular, we will construct equations whose left-hand-side is
obtained integrating the probability density function of the
eigenvalue distribution, and the right-hand-side by exploiting
identities based on the traces of the powers of a matrix.

We write $\mu'_z$ for the $z^{th}$ raw moment (moment about the
origin) of the eigenvalue distribution, and $\mu_z$ for the
corresponding central moment (moment about the mean). For any matrix
with zero on the diagonal (and hence $\overline{\lambda} = 0$), the
raw and central moment are the same:

\begin{equation}
  \mu'_z = \mu_z = \frac{1}{s} \sum_{j = 1}^s \lambda_j^z
  = \frac{\Tr(M^z)}{s}
\end{equation}
Because we are dealing with real matrices, all eigenvalues are either
real or complex conjugates. It follows that all the moments $\mu'_z$
are real. Moreover, this greatly simplifies the integrals for the
moments. We write $\lambda_j = x_j + iy_j$, where $i = \sqrt{-1}$. The
integral for the second moment is:

\begin{equation}
  \mu'_2 = \mu_2 = \int \lambda^2 P(\lambda) d\lambda = \iint (x + iy)^2 P(x,y) dx dy =
  \iint (x^2 - y^2 + 2ixy) P(x,y) dx dy
\end{equation}
\noindent where $P( \cdot)$ is the probability density
function. Because the complex eigenvalues are paired, if $x + iy$ is
an eigenvalue, so is $x -iy$. Hence, the imaginary part vanishes from
the integral, leaving

\begin{equation}
  \mu'_2 = \mu_2  = \iint (x^2 - y^2) P(x,y) dx dy
\end{equation}
Similarly, we can write

\begin{equation}
  \mu'_4 = \mu_4  = \iint (x^4 + y^4 - 6 x^2 y^2) P(x,y) dx dy
\end{equation}

\subsection*{Semi-superelliptical distribution.}

As in the main text, we start by setting $\rho = 1$, so that $M$ is
symmetric (Hermitian). In this case, all eigenvalues are real,
simplifying the problem. Whenever $\rho = -1$, $M$ is
skew-symmetric, and all eigenvalues are imaginary.

For $\rho =1$ and $k \to \infty$, the eigenvalue distribution is
described by Wigner's semicircle law:

\begin{equation}
  \Pr(\lambda = x) = P(x) =  \frac{2 \sqrt{(2r)^2 - x^2}}{(2r)^2 \pi}
\end{equation}
\noindent where we write the radius as $a = 2r$ to make the derivation
consistent with the asymmetric case below. The distribution of the
eigenvalues has mean 0, and variance:

\begin{equation}
  \mu_2(r) = \int_{-2r}^{2r} x^2 P(x) dx = r^2
\end{equation}
Because, for these matrices

\begin{equation}
   \mu_2(r) = \frac{\Tr(M^2)}{s} = \mathbb E [M_{ij}M_{ji}] = \mathbb
   E [M_{ij}^2] = 1
\end{equation}
\noindent then necessarily $r = 1$.

We want to extend Wigner's semicircle law to the case of matrices with
complex network structure. We hypothesize that in this case, the
density of the eigenvalues follows a semi-superellipse:

\begin{equation}
  \Pr(\lambda = x)  = P(x)= \frac{2 \sqrt[n]{(2r)^n - x^n}}{4 (2r)^2 \Gamma(1 +
    \sfrac{1}{n})^2 \Gamma(1 + \sfrac{2}{n})^{-1}}
\end{equation}
Because we have two parameters, $r$ and $n$, we need to compute two
moments. The first two nonzero moments for this distribution are
$\mu_2(r,n)$ and $\mu_4(r,n)$.

\begin{equation}
  \mu_2(r,n) = \int_{-2r}^{2r} x^2 P(x) dx =
  r^2 \frac{2 \Gamma\left( \frac{2}{n} \right) \Gamma\left( \frac{3}{n} \right)}
  {\Gamma\left( \frac{1}{n} \right) \Gamma\left( \frac{4}{n} \right)}  
\end{equation}
Given that also in this case we have $\mu_2(r,n) = 1$, we can solve
for the positive $r$:

\begin{equation}
  r
  = \sqrt{\frac{\Gamma\left( \frac{1}{n} \right) \Gamma\left( \frac{4}{n} \right)}{
  2 \Gamma\left( \frac{2}{n} \right) \Gamma\left( \frac{3}{n} \right)}}
  \label{myr}
\end{equation}
Integrating to obtain $\mu_4(r,n)$:

\begin{equation}
  \mu_4(r,n) = \int_{-2r}^{2r} x^4 P(x) dx = \frac{8 r^4 \Gamma
    \left(\frac{1}{n}\right) \Gamma \left(1 + \frac{2}{n}\right) \Gamma
    \left(1 + \frac{5}{n}\right)} {15 \Gamma
    \left(1+\frac{1}{n}\right)^2 \Gamma \left(\frac{6}{n}\right)}
\end{equation}
\noindent and substituting the value of $r$:

\begin{equation}
 \mu_4(n) = \frac{
                 2 
                 \Gamma\left( \frac{1}{n} \right)^3
                 \Gamma\left( \frac{4}{n} \right)^2
                 \Gamma\left( 1 + \frac{2}{n} \right)
                 \Gamma\left( 1 + \frac{5}{n} \right)
                 }{
                 15
                 \Gamma\left( 1 + \frac{1}{n} \right)^2
                 \Gamma\left( \frac{2}{n} \right)^2
                 \Gamma\left( \frac{3}{n} \right)^2
                 \Gamma\left( \frac{6}{n} \right)
                 }
\end{equation}
Hence, knowing the value of $\mu_4(n) = \Tr(M^4)/s$, we can
numerically solve the equation above for $n$, and recover $a = 2 r$
from $n$ using Eq. (\ref{myr}).

\subsection*{$\boldsymbol k$-regular graphs: superelliptical distribution.}

We now derive the superelliptical distribution for eigenvalues that
are uniformly distributed in a superellipse. Our simulations show that
this is the case for matrices whose underlying structure is a
$k$-regular random graph. For those generated using Erd\H{o}s-R\'enyi
random graphs, the distribution slightly departs from uniformity, and
for those built using Power-law graphs, it departs severely. Hence,
only for the case of $k$-regular graphs do we expect the
superelliptical uniform distribution to hold.

We make only one assumption in order to derive the values of $n$ and
$r$. We assume that $a + b = 2 r$ irrespective of $\rho$. This
assumption is strongly supported by numerical simulations for the case
of $k$-regular random graphs (the only case in which we believe the
following results to hold). When $a + b = 2 r$ for any $\rho$, the
semi-superelliptical distribution derived above is recovered for $\rho
= 1$. As such, the value of $r$ is that in Eq. (\ref{myr}).

As we did above, we first compute the moments using the eigenvalue
distribution, and then derive the right-hand-sides of the equations
using traces.

The probability density function for eigenvalues uniformly distributed
in the superellipse $\sfrac{|x|^n}{a^n} + \sfrac{|y|^n}{b^n} \leq 1$
is:

\begin{equation}
  P(x,y) = \frac{1}{4 a b \Gamma(1 +
    \sfrac{1}{n})^2 \Gamma(1 + \sfrac{2}{n})^{-1}}
\end{equation}
Integrating $(x^2 - y^2)$ we obtain $\mu_2(a,b,n)$:

\begin{equation}
  \mu_2(a,b,n) = \int_{-a}^{a} \int_{-\frac{b}{a} \sqrt[n]{a^n -
      |x|^n}}^{\frac{b}{a} \sqrt[n]{a^n - |x|^n}} (x^2 - y^2) P(x,y)
  dy dx = \frac{(a-b) (a+b) \Gamma \left(\frac{2}{n}\right) \Gamma
    \left(\frac{3}{n}\right)}{2 \Gamma \left(\frac{1}{n}\right) \Gamma
    \left(\frac{4}{n}\right)}
\end{equation}

Because $\mu_2(a,b,n) = \mathbb E[M_{ij}M_{ji}] = \rho$, we can set $b
 = 2r - a$ and solve for $a$:

\begin{equation}
  a = r
  + \frac{\rho \Gamma \left(\frac{1}{n}\right) \Gamma \left(\frac{4}{n}\right)}{
  2 r \Gamma \left(\frac{2}{n}\right) \Gamma \left(\frac{3}{n}\right)}
  = r(1 + \rho)
\end{equation}

Similarly, $b = r (1 -\rho)$, consistently with the fact that,when
$k \to \infty$, then $a = 1 + \rho$ and $b = 1 - \rho$ (elliptical
law\cite{girko1985,sommers1988spectrum,Naumov20xx}). 

We then move to the fourth moment:

\begin{equation}
\begin{aligned}
  \mu_4(a,b,n) =& \int_{-a}^{a} \int_{-\frac{b}{a} \sqrt[n]{a^n -
      |x|^n}}^{\frac{b}{a} \sqrt[n]{a^n - |x|^n}} (x^4 + y^4 - 6 x^2
  y^2 ) P(x,y) dy dx =\\ 
  =& 
  \frac{a^4 \Gamma\left( \frac{1}{n} \right) \Gamma\left(1
  + \frac{2}{n} \right) \Gamma\left( 1
  + \frac{5}{n} \right)}{30 \Gamma\left(1 + \frac{1}{n} \right)^2 \Gamma\left( \frac{6  }{n} \right)} + \\
  & \frac{b^4 \Gamma\left( 1 + \frac{2}{n} \right) \Gamma\left( 1
  + \frac{5}{n} \right)}{5 \Gamma\left( 1
  + \frac{1}{n} \right) \Gamma\left( 1 + \frac{6}{n} \right)} - \\
  & \frac{a^2 b^2 2^{1-\frac{4}{n}} \Gamma\left( \frac{1}{2}
  + \frac{1}{n} \right) \Gamma\left( \frac{3}{n} \right)}{30 \Gamma\left(1 + \frac{1}{n} \right)^2 \Gamma\left( \frac{6}{n} \right)}
\end{aligned}
\end{equation}
Substituting $a = r(1 + \rho)$, $b = r(1-\rho)$ and replacing $r$ with
the value in Eq. (\ref{myr}) greatly simplifies the expression:

\begin{equation}
  \mu_4(n,\rho) = \frac{2^{\frac{4}{n}-2} 
  \Gamma \left(\frac{1}{2}+\frac{2}{n}\right) 
  \Gamma \left(\frac{4}{n}\right) 
  \left(\frac{\left({\rho}^4+6 {\rho}^2+1\right) 
  \Gamma \left(\frac{1}{n}\right) 
  \Gamma \left(\frac{5}{n}\right)}
  {\Gamma \left(\frac{3}{n}\right)^2}-
  3 \left(1 - {\rho}^2\right)^2\right)}{3 \sqrt{\pi } 
  \Gamma \left(\frac{6}{n}\right)}
\end{equation}
Knowing $\rho$ and $\Tr(M^4)$, we solve this equation numerically to
obtain $n$, use Eq. (\ref{myr}) to obtain $r$, and, finally, use $r$
and $\rho$ to compute $a$ and $b$.

\subsection*{Computing expected traces.}
Above, we computed the values of $n$, $a$ and $b$ using the actual,
observed traces of $M^2$ and $M^4$: $\mu_2 = \Tr(M^2)/s$ and
$\mu_4=\Tr(M^4)/s$. Alternatively, one can compute the expectations
for the traces when the distribution of the coefficients is known and
one can count certain structures in the graph. Clearly $\mathbb
E[\Tr(M^2)/s] = \rho$. The computation of $\mathbb E[\Tr(M^4)/s]$ is
detailed below for the case of symmetric and asymmetric
matrices. Because the results are particularly simple for matrices
with $k$-regular graph structure, we explore the effect of the average
degree on the parameters of the superellipses in this case.

\paragraph*{Symmetric matrices.}
Only four possible network structures contribute to the fourth power
of $M$ in the graphs analyzed here. In fact, the diagonal element
$M_{jj}^4$ is composed of the sum of the closed paths involving four
links that start and end at node $j$. There are four possibilities
(Table S2, all indices are taken to be different -- $j \neq l \neq n
\neq m$):
\begin{itemize}
  \item[T1)] $M_{jl}M_{lj}M_{jl}M_{lj}$, i.e., the same undirected
    link tread twice. For symmetric matrices, we can write $M_{jl}^4$.
  \item[T2)] $M_{jl}M_{lj}M_{jm}M_{mj}$, i.e., a path where we move
    from $j$ to $l$, come back and then move to $m$ and come back. For
    symmetric matrices, we can write $M_{jl}^2 M_{jm}^2$.
  \item[T3)] $M_{jl}M_{lm}M_{ml}M_{lj}$, i.e., a path where we move
    from $j$ to $l$, from $l$ to $m$ and then go back to $l$ and
    finally to $j$. For symmetric matrices, we can write $M_{jl}^2
    M_{lm}^2$.
  \item[T4)] $M_{jl}M_{lm}M_{mn}M_{nj}$, i.e., a four-link directed
    cycle.
\end{itemize}

For each graph, we can count how many ways there are to obtain each
structure:

\begin{itemize}
  \item[T1)] There is an occurrence of T1 for each connection in the
    graph. Thus, the total number of structures of this kind is $s k$.
  \item[T2)] For a node of degree $k_j$, there are $\binom{k_j}{2}$
    ways of choosing the two partners. However, for each choice of
    partners ($l$ and $m$) we have two possibilities: visit $l$ first,
    and visit $m$ first. Hence, the number of structures is: $2
    \sum_{j=1}^{s} \binom{k_j}{2}$.
  \item[T3)] For each edge connecting $j$ to $l$, we have $k_l -1$
    ways of choosing the third partner. Thus, the number of structures
    is $\sum_{j=1}^{s} \sum_{l=1}^{s} A_{jl}(k_l - 1)$.
  \item[T4)] The number of four-cycles in the graph cannot be
    expressed as a simple function of the degrees of the nodes.
\end{itemize}

Finally, we need to compute the expectation for their values. For the
normal bivariate distribution described above, we have:

\begin{enumerate}
\item[T1)] $\mathbb E[M_{jl}M_{lj}M_{jl}M_{lj}] = \mathbb E[M_{jl}^4]
  = \frac{3}{k^2}$, as the fourth central moment of a normal
  distribution $\mathcal N(0, \sigma^2)$ is simply $3 \sigma^2$.
\item[T2)] $\mathbb E[M_{jl}M_{lj}M_{jm}M_{mj}] = \mathbb E[M_{jl}^2
  M_{jm}^2] = \frac{1}{k^2}$ (the product of the two variances).
\item[T3)] $\mathbb E[M_{jl}M_{lm}M_{ml}M_{lj}] = \mathbb E[M_{jl}^2
    M_{lm}^2] = \frac{1}{k^2}$.
\item[T4)] $\mathbb E[M_{jl}M_{lm}M_{mn}M_{nj}] = 0$, as the
  expectation of the product is simply the product of the expectations
  -- all the coefficients are independent.
\end{enumerate}

Given that $\mathbb E[T4] = 0$, summing $(\# T1 \cdot \mathbb E[T1] +
\# T2 \cdot \mathbb E[T2] + \# T3 \cdot \mathbb E[T3])$ we obtain
$\mathbb E[Tr(M^4)]$. Dividing by $s$, we obtain $\mathbb E[Tr(M^4)/s]
= \mu_4(n)$. The calculation is especially simple for a $k$-regular
graph, in which we can count the number of T2 and T3 structures very
easily. For this type of graph, $\mathbb E[Tr(M^4)/s] = 2
+ \frac{1}{k}$.

\setcounter{table}{0}
\begin{table}
  \begin{tabular}{zzzzz}
    \hline
    Type & Graph & Coefficients & Number of Occurrences & Expectation\\
    \hline
    T1 &
    \begin{tikzpicture}[scale=0.35]
      \Vertex{j}
      \Vertex[x=3,y=0]{l}
      \Edge[style={->,bend left}](j)(l)
      \Edge[style={->,bend left=60}](j)(l)
      \Edge[style={->,bend left}](l)(j)
      \Edge[style={->,bend left=60}](l)(j)
    \end{tikzpicture}
    &
    $M_{jl}M_{lj}M_{jl}M_{lj} = M_{jl}^4$
    &
    $s k$
    &
    $\frac{3}{k^2}$\\
    
    T2 &
    \begin{tikzpicture}[scale=0.35]
      \Vertex{j}
      \Vertex[x=2,y=2]{l}
      \Vertex[x=2,y=-2]{m}
      \Edge[style={->,bend left}](j)(l)
      \Edge[style={->,bend left}](j)(m)
      \Edge[style={->,bend left}](l)(j)
      \Edge[style={->,bend left}](m)(j)
    \end{tikzpicture}
    &
    $M_{jl}M_{lj}M_{jm}M_{mj} = M_{jl}^2 M_{jm}^2$
    &
    $2 \sum_{j=1}^{s} \binom{k_j}{2}$
    &
    $\frac{1}{k^2}$\\

    T3 &
    \begin{tikzpicture}[scale=0.35]
      \Vertex{j}
      \Vertex[x=3,y=0]{l}
      \Vertex[x=6,y=0]{m}
      \Edge[style={->,bend left}](j)(l)
      \Edge[style={->,bend left}](l)(m)
      \Edge[style={->,bend left}](m)(l)
      \Edge[style={->,bend left}](l)(j)
    \end{tikzpicture}
    &
    $M_{jl}M_{lm}M_{ml}M_{lj} = M_{jl}^2 M_{lm}^2$
    &
    $\sum_{j=1}^{s} \sum_{l=1}^{s} A_{jl}(k_l - 1)$
    &
    $\frac{1}{k^2}$\\

    T4 &
    \begin{tikzpicture}[scale=0.35]
      \Vertex{j}
      \Vertex[x=3,y=0]{l}
      \Vertex[x=3,y=3]{m}
      \Vertex[x=0,y=3]{n}
      \Edge[style={->,bend right}](j)(l)
      \Edge[style={->,bend right}](l)(m)
      \Edge[style={->,bend right}](m)(n)
      \Edge[style={->,bend right}](n)(j)
    \end{tikzpicture}
    &
    $M_{jl}M_{lm}M_{mn}M_{nj}$
    &
    Depends on graph
    &
    $0$\\
    \hline
  \end{tabular}
  \caption{\textit{The structures contributing to $\Tr(M^4)$. For each
      structure, we report the form of the coefficients, the number of
      occurrences in a graph, and the expected value when the
      coefficients are taken from a bivariate normal distribution with
      correlation $\rho =1$.}}
\end{table}

\paragraph*{Asymmetric matrices.}

As in the case of symmetric matrices, the expectation $\mathbb
E[\Tr(M^4)/s]$ can be obtained counting the structures T1, T2 and T3
in the graph. However, we need to re-compute the expectations for the
structures for arbitrary $\rho$ (Table S3). For a $k$-regular random
graph, we have $\mathbb E[\Tr(M^4)/s] = 2 \rho^2 + \frac{1}{k}$.

\begin{table}
  \begin{tabular}{zzzzz}
    \hline
    Type & Graph & Coefficients & Number of Occurrences & Expectation\\
    \hline
    T1 &
    \begin{tikzpicture}[scale=0.35]
      \Vertex{j}
      \Vertex[x=3,y=0]{l}
      \Edge[style={->,bend left}](j)(l)
      \Edge[style={->,bend left=60}](j)(l)
      \Edge[style={->,bend left}](l)(j)
      \Edge[style={->,bend left=60}](l)(j)
    \end{tikzpicture}
    &
    $M_{jl}M_{lj}M_{jl}M_{lj} = M_{jl}^2 M_{lj}^2$
    &
    $s k$
    &
    $\frac{1 + 2 \rho^2}{k^2}$\\
    
    T2 &
    \begin{tikzpicture}[scale=0.35]
      \Vertex{j}
      \Vertex[x=2,y=2]{l}
      \Vertex[x=2,y=-2]{m}
      \Edge[style={->,bend left}](j)(l)
      \Edge[style={->,bend left}](j)(m)
      \Edge[style={->,bend left}](l)(j)
      \Edge[style={->,bend left}](m)(j)
    \end{tikzpicture}
    &
    $M_{jl}M_{lj}M_{jm}M_{mj}$
    &
    $2 \sum_{j=1}^{s} \binom{k_j}{2}$
    &
    $\frac{\rho^2}{k^2}$\\

    T3 &
    \begin{tikzpicture}[scale=0.35]
      \Vertex{j}
      \Vertex[x=3,y=0]{l}
      \Vertex[x=6,y=0]{m}
      \Edge[style={->,bend left}](j)(l)
      \Edge[style={->,bend left}](l)(m)
      \Edge[style={->,bend left}](m)(l)
      \Edge[style={->,bend left}](l)(j)
    \end{tikzpicture}
    &
    $M_{jl}M_{lm}M_{ml}M_{lj}$
    &
    $\sum_{j=1}^{s} \sum_{l=1}^{s} A_{jl}(k_l - 1)$
    &
    $\frac{\rho^2}{k^2}$\\

    T4 &
    \begin{tikzpicture}[scale=0.35]
      \Vertex{j}
      \Vertex[x=3,y=0]{l}
      \Vertex[x=3,y=3]{m}
      \Vertex[x=0,y=3]{n}
      \Edge[style={->,bend right}](j)(l)
      \Edge[style={->,bend right}](l)(m)
      \Edge[style={->,bend right}](m)(n)
      \Edge[style={->,bend right}](n)(j)
    \end{tikzpicture}
    &
    $M_{jl}M_{lm}M_{mn}M_{nj}$
    &
    Depends on graph
    &
    $0$\\
    \hline
  \end{tabular}
  \caption{\textit{The structures contributing to $\Tr(M^4)$. For each
      structure, we report the form of the coefficients, the number of
      occurrences in a graph, and the expected value when the
      coefficients are taken from a bivariate normal distribution with
      correlation $\rho$.}}
\end{table}

\begin{figure}
\begin{centering}
  \includegraphics[width = 0.65\linewidth]{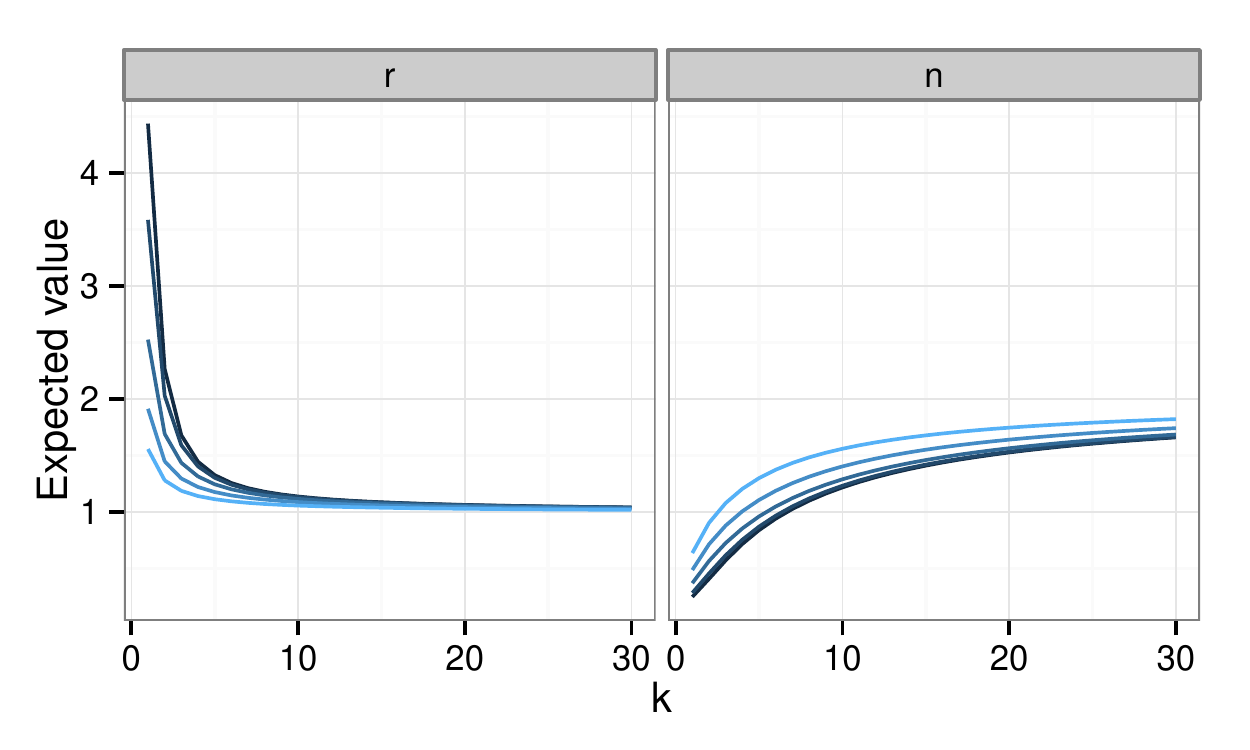}
  \caption{\textit{Expected $r$ and $n$ for matrices whose underlying
  structure is a $k$-regular graph of degree $k$ (x-axis). Colors
  represent the value of $\rho$ (from 0 -- darker color, to 1).}}
  \label{kreg}
  \end{centering}
\end{figure}

\paragraph*{$\boldsymbol k$-regular graphs.} 
As shown in the two sections above, for a matrix $M$ with $k$-regular
graph structure, we have that $\mathbb E[\Tr(M^2)/s] = \rho$, and
$\mathbb E[\Tr(M^4)/s] = 2 \rho^2 + \frac{1}{k}$. As such, we can
easily find the expected $n$ and $r$ for any value of $\rho$ and
$k$. We find that, when increasing $k$, $r$ converges to 1 quite
rapidly, while the value of $n$ converges to 2 more slowly. Because
the value of $r$ depends only on $n$, and the value of $n$ depends on
the average degree $k$ and on $\rho^2$, without loss of generality we
can assume $\rho$ to be positive. The results show that a strong
correlation (large $\rho$) speeds up the convergence of both $r$ and
$n$ (Figure~\ref{kreg}).

\subsection*{Supplementary Results.}
\paragraph*{Normal bivariate distribution, negative correlation.}

\begin{sidewaysfigure}
   \includegraphics[width
    = \linewidth]{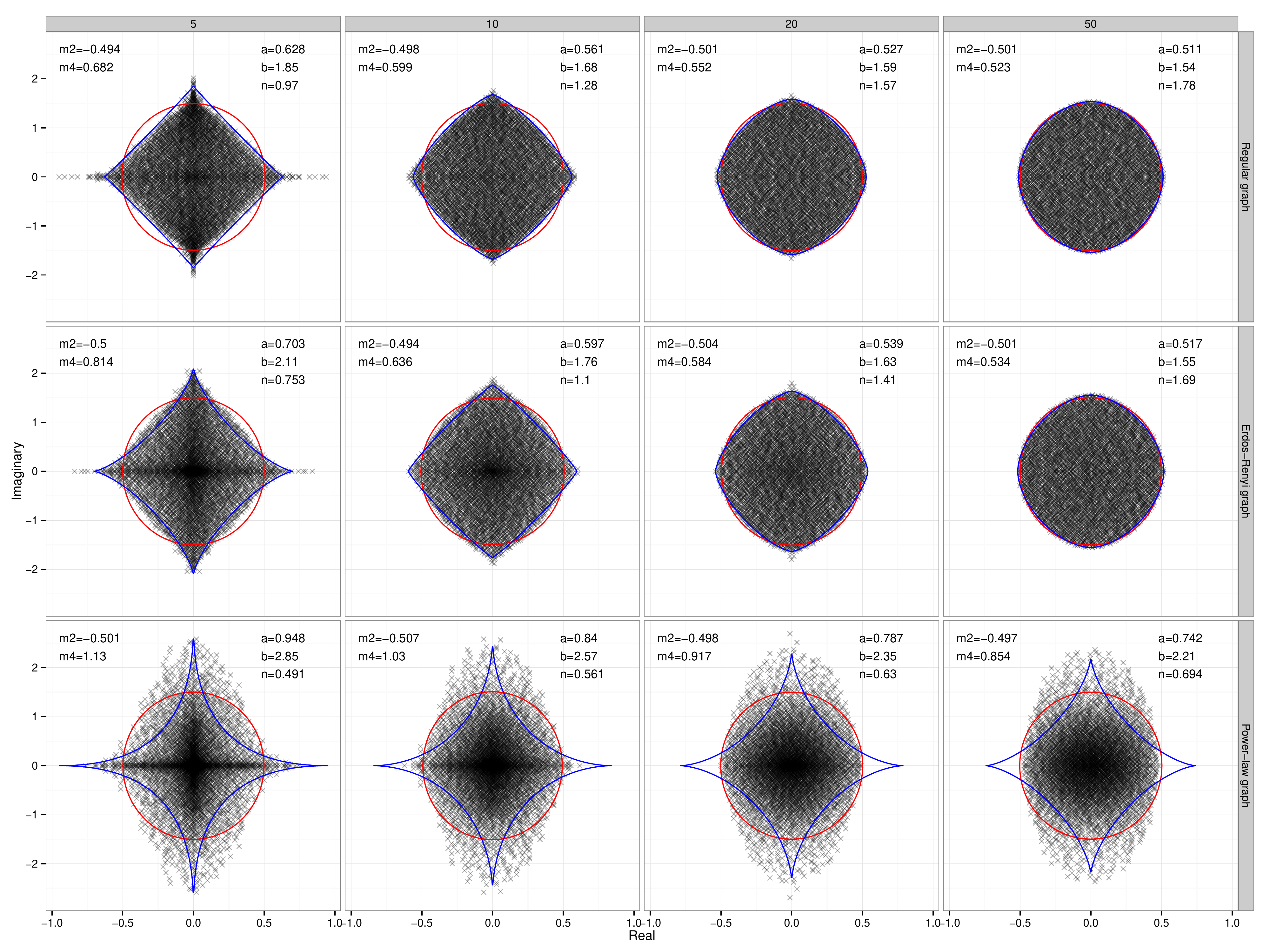} \caption{\textit{As
    main text Figure 2 and Figure 3, but with negative correlation:
    $\mathbb E[\rho] = -0.5$.}}  \label{NegCorr}
\end{sidewaysfigure}

In the main text, we drew the eigenvalue distributions for asymmetric
matrices with complex network structure when $\mathbb E[\rho] = 0$
(Figure 2) and $\mathbb E[\rho] = 0.5$ (Figure 3). In
Figure \ref{NegCorr}, we show the case of $\mathbb E[\rho] = - 0.5$.

\paragraph*{Uniform distribution.}
Girko's circular law has been recently shown to be universal, i.e.,
the law holds given very mild conditions on the distribution of the
coefficients in the matrix\cite{tao2010random}. Hence, it is
interesting to test whether choosing different distributions for the
coefficients in the matrix significantly alters the superelliptical
distribution. As with Girko's circular law, we find that our results
hold for different distributions.

\begin{sidewaysfigure}
  \includegraphics[width = \linewidth]{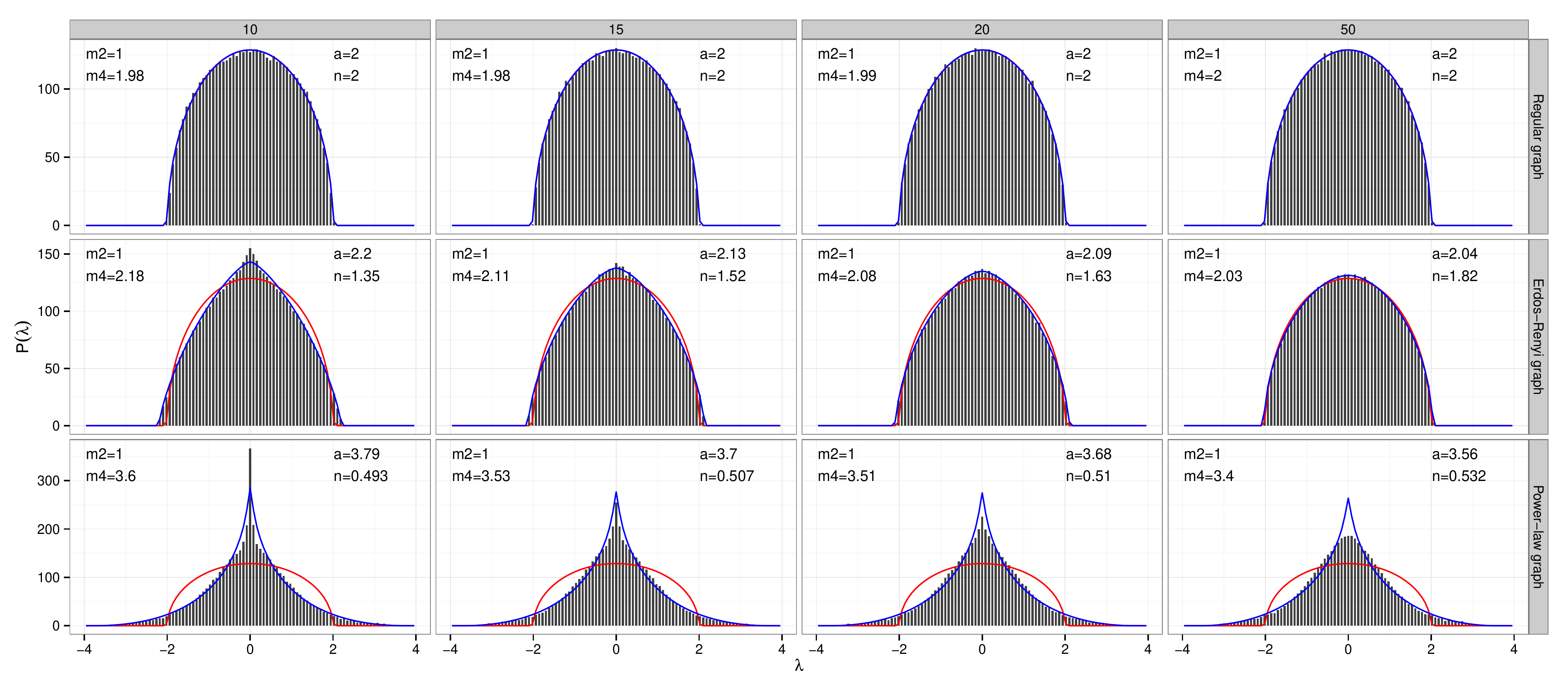}
  \caption{\textit{As main text Figure 1, but with entries $M_{ij} =
      M_{ji} = A_{ij} U_{ij}$, where $U$ is a symmetric matrix with
      uniformly distributed coefficients (see text). }}
  \label{UniSymm}
\end{sidewaysfigure}

For example, take symmetric matrices whose coefficients are sampled
from a uniform distribution. We set $M_{ij}=M_{ji}=A_{ij}U_{ij}$,
where $U$ is a symmetric matrix whose coefficients come from $\mathcal
U(-\sqrt{\sfrac{3}{k}}, \sqrt{\sfrac{3}{k}})$, so that $\mathbb E
[M_{ij}] = 0$, and $\text{Var}[M_{ij}] = \mathbb E [M_{ij}^2] =
\sfrac{1}{s}$. The eigenvalue density follows the semi-superelliptical
distribution, as shown in Figure \ref{UniSymm}.

\begin{sidewaysfigure}
  \includegraphics[width = \linewidth]{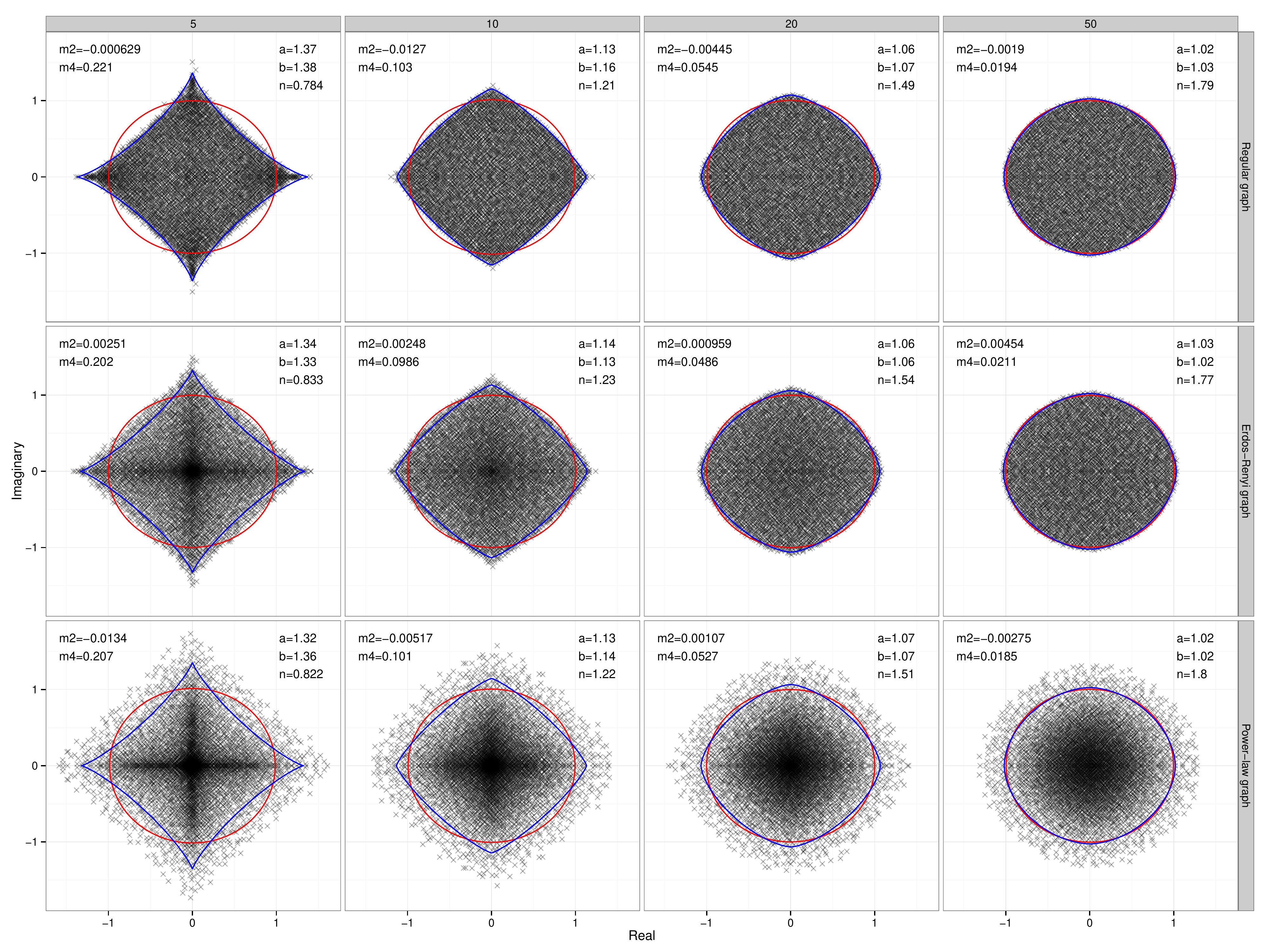}
  \caption{\textit{As main text Figure 2, but with entries $M_{ij} =
      A_{ij} U_{ij}$, where $U$ is a matrix whose coefficients are
      uniformly distributed.}}
  \label{UniAsymm}
\end{sidewaysfigure}

Similarly, make the coefficients in $U$ independent identically
distributed: then matrices whose underlying structure are $k$-regular
graphs follow the superelliptical distribution
(Figure~\ref{UniAsymm}).

\section*{Approximating the largest eigenvalue of adjacency matrices}

In the analysis above, we assumed the mean of the coefficients of $M$
to be $0$. The problem is more complicated when this is not the
case. However, many applications deal with non-negative matrices
(i.e., whose entries are $\geq 0$), and therefore it is important to
extend the methods above to encompass such cases.

In the following paragraphs, we deal with adjacency matrices of
undirected graphs with complex network structure. This case is
particularly well-studied, given the potential for applications, and
many bounds on the largest eigenvalue (spectral radius) have been
proposed (for a brief survey, see Das \& Kumar\cite{das2004some}). One
of the simplest and most beautiful approximations for the largest
eigenvalue is that proposed by Chung \textit{et
al.}\cite{chung2003spectra}: $\lambda_1 \approx \overline{k^2}
/ \overline{k}$ (the average of the squared degrees divided by the
average degree). This approximation---which was also derived using a
completely different approach by Nadakuditi \&
Newman\cite{nadakuditi2013spectra}---is expected to hold whenever the
minimum degree is large enough compared to the average degree:
$\min(k_i) \gg \sqrt{\overline{k}} \log^3 s$.

The spectrum of the adjacency matrix of a complex network generally
departs severely from Wigner's
semicircle\cite{farkas2001spectra}. Moreover, although the ``bulk'' of
the eigenvalues falls in a well-defined region, the largest
eigenvalue, $\lambda_1$---and, possibly, other large eigenvalues---is
not localized in this region.

\begin{figure}
  \includegraphics[width
  = \linewidth]{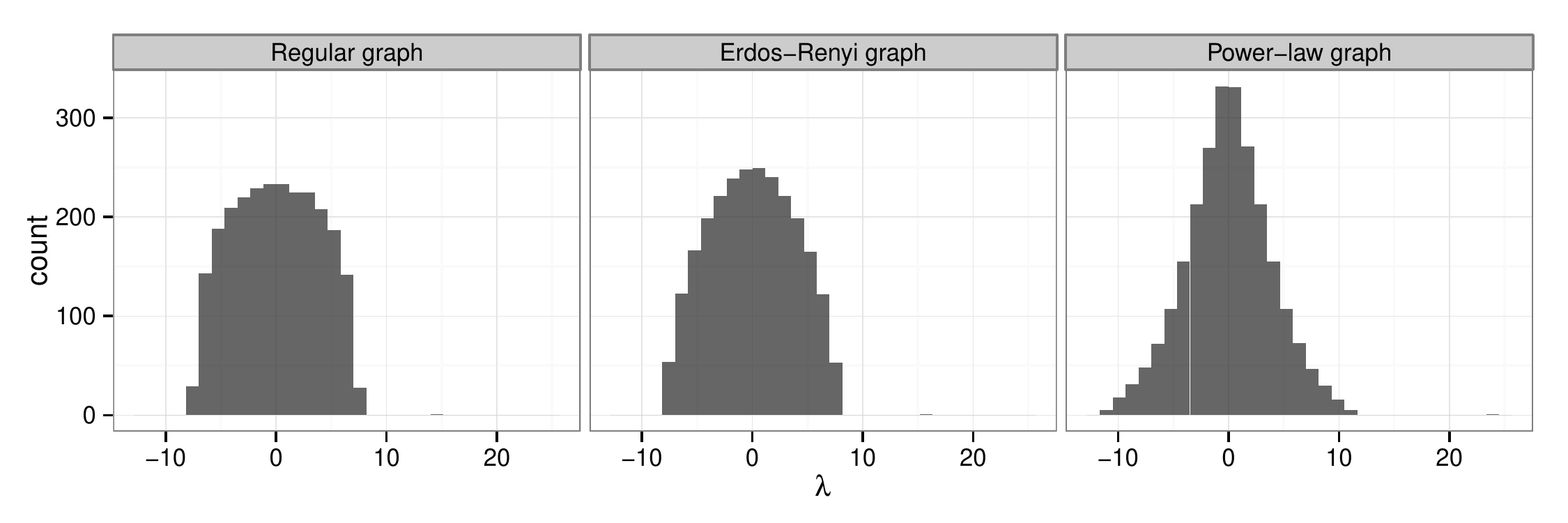} \caption{\textit{Spectra
  of adjacency matrices of size $2500$ built using the configuration
  model with average degree $k=15$, and degree distributions specified
  by different models. In all cases, the spectrum can be approximated
  by a semi-superellipse containing all the eigenvalues but the
  dominant one, plus $\lambda_1$ itself.}}  \label{AdjSpec}
\end{figure}

Here we present a novel method to estimate the value of $\lambda_1$
obtained assuming that all the other eigenvalues fall in a symmetric
distribution, such as a semi-superellipse. This seems to be the case
for adjacency matrices of random graphs with arbitrary degree
distributions built using the configuration model
(Figure~\ref{AdjSpec}).

We consider the adjacency matrix $A$, describing a simple, undirected
graph without self-loops. In this matrix, the diagonal is $0$, and
therefore the mean of the eigenvalues is zero. We can write:

\begin{equation}
  \Tr(A) = 0 = \sum_{i=1}^s \lambda_i = \lambda_1 + \sum_{i=2}^s
  \lambda_i
\end{equation}
\noindent where $\lambda_1$ is the largest eigenvalue (spectral
radius). We want to describe the distribution of the non-dominant
eigenvalues $\lambda_2, \ldots, \lambda_s$. In general, we can write:

\begin{equation}
  \Tr(A^j) = \sum_{i=1}^s \lambda^j_i = s \mu'_j 
\end{equation}
That is, the trace of the $j^{th}$ power of $M$ is equal to the size
$s$ times the $j^{th}$ raw moment of the eigenvalue distribution. When
we consider $\lambda_1$ separately, we have:

\begin{equation}
  \Tr(A^j) = s \mu'_j = (s-1) \tilde{\mu'}_j + \lambda_1^j
\end{equation}
\noindent where with $\tilde{\mu'}_j$ we indicate the $j^{th}$ raw
moment of the distribution of the eigenvalues when we exclude the
dominant one. Because the trace of the matrix is zero, we have:

\begin{equation}
  \begin{aligned}
    \Tr(A) = 0 = (s-1) \tilde{\mu'}_1 + \lambda_1 & \to &
    \tilde{\mu'}_1 = - \frac{\lambda_1}{s-1}
  \end{aligned}
\end{equation}

The trace of $A^2$ is simply the number of connections:
\begin{equation}
  \begin{aligned}
    \Tr(A^2) = s k = (s-1) \tilde{\mu'}_2 + \lambda_1^2 & \to &
    \tilde{\mu'}_2 = \frac{s k- \lambda_1^2}{s-1}
  \end{aligned}
\end{equation}

In general, the $j^{th}$ raw moment is:
\begin{equation}
  \tilde{\mu'}_j = \frac{\Tr(A^j) - \lambda_1^j}{s-1}
\end{equation}

If the distribution of all the eigenvalues but $\lambda_1$ is
approximately symmetric, then the odd central moments $\tilde{\mu}_{2j
+1} \approx 0$. We can exploit this fact to approximate the value of
$\lambda_1$. First, we need the following formulas relating raw and
central moments:

\begin{equation}
  \begin{aligned} \mu'_3 =& \mu_3 + 3 \mu'_1\mu'_2 - 2
    (\mu'_1)^3\\ 
    \mu'_5 =& \mu_5 + 5 \mu'_1\mu'_4 - 10
    (\mu'_1)^2\mu'_3 + 10 (\mu'_1)^3\mu'_2 - 4
    (\mu'_1)^5\\ 
\end{aligned} 
\label{rawcentral}
\end{equation}

Then, we can for example, choose $\tilde{\mu}_{3} \approx 0$. In this
case,
\begin{equation}
    \Tr(A^3) = (s-1) \tilde{\mu'}_3 + \lambda_1^3  \approx 
   (s-1)\left(
   3 \tilde{\mu'}_1 \tilde{\mu'}_2 - 2 (\tilde{\mu'}_1)^3 
   \right)+ \lambda_1^3    
\end{equation}

Substituting the values for $\tilde{\mu'}_1$ and $\tilde{\mu'}_2$, we
obtain:
\begin{equation}
   \lambda_1^3 \frac{s (s + 1)}{(s-1)^2} - \lambda_1 \frac{3 k
   s}{s-1} \approx \Tr(A^3)
\end{equation}

\noindent which, assuming equality, we can solve numerically, or
   analytically (taking the only real value):

\begin{equation}
\footnotesize
\lambda_1 \approx 
\frac{-\sqrt[3]{2} \left(s^2 \left(s^2-1\right) 
\left(\sqrt{\left(s^2-1\right) 
\left(\left(s^2-1\right) 
\Tr(A^3)^2-4 k^3 s^2\right)}
+s^2 (-\Tr(A^3))+\Tr(A^3)\right)\right)^{2/3}
-2 k s^2 \left(s^2-1\right)}{2^{2/3} s (s+1) 
\sqrt[3]{s^2 \left(s^2-1\right) 
\left(\sqrt{\left(s^2-1\right) \left(\left(s^2-1\right) 
\Tr(A^3)^2-4 k^3 s^2\right)}
+s^2 (-\Tr(A^3))+\Tr(A^3)\right)}}
\end{equation}

Similarly, we can choose to zero a higher moment. For example, if
$\tilde{\mu}_{5} \approx 0$, we have:

\begin{equation}
    \Tr(A^5) = (s-1) \tilde{\mu'}_5 + \lambda_1^5  \approx 
   (s-1)\left(
   5 \tilde{\mu'_1}\tilde{\mu'}_4 - 10
    (\tilde{\mu'}_1)^2\tilde{\mu'}_3 + 10 (\tilde{\mu'}_1)^3\tilde{\mu'}_2 - 4
    (\tilde{\mu'}_1)^5  \right)+ \lambda_1^5
\end{equation}

\noindent which yields
\begin{equation}
\lambda_1^5 \frac{s (s + 1) (s^2 + 1)}{(s-1)^4} 
- \lambda_1^3 \frac{10 s k}{(s-1)^3} 
- \lambda_1^2 \frac{10 \Tr(A^3)}{(s-1)^2} 
- \lambda_1 \frac{5 \Tr(A^4)}{(s-1)} \approx \Tr(A^5)
\end{equation}

Assuming equality, one can solve the equation numerically. Clearly,
one can derive an approximation assuming any of the odd central
moments $\tilde{\mu}_{2j + 1}$ to be zero.

\section*{Beyond the configuration model: the spectra of real networks}

In the previous sections, we dealt with matrices whose underlying
structure was built using the configuration model. That is, the
networks contain no special structure, besides the imposition of a
degree distribution. They are, in a way, ``random networks''. Here we
explore the sensitivity of our results to the use of networks that
depart considerably from this assumption. In fact, the ultimate test
for any of these methods is to be able to deal with real-world
networks.

We repeat the analysis above using three different data sets: i)
networks built according to the Watts-Strogatz
model\cite{watts1998collective}; ii) networks built with the
Barab\'asi-Albert model\cite{barabasi1999emergence}; and, iii) A set
of four networks of biological interest, describing a) the structure
of the contact network in an high school\cite{salathe2010high}, b) the
neural network of the worm
\textit{C. elegans}\cite{White12111986,watts1998collective}, c) the
metabolic network of \textit{C. elegans}\cite{duch2005community} and,
d) the food web structure of the Weddell Sea
ecosystem\cite{jacob2005trophic}. For simplicity, all the networks are
taken to be undirected.

For the networks built using the Watts-Strogatz model, we set $s =
5000$, and arranged the nodes in a one-dimensional lattice without
boundaries. Each node is connected to the $\sfrac{k}{2}$ nodes on its
left and to its right, forming a regular, one-dimensional
lattice. Links are then rewired with probability $0.05$. This can be
accomplished calling the routine~\texttt{watts.strogatz.game(1, 5000,
k/2, 0.05)} in~\texttt{igraph}.

For the networks built using the Barab\'asi-Albert model, we built
networks of size $s =5000$ using preferential attachment with exponent
$1$, where each node introduced in the network attaches to
$\sfrac{k}{2}$ nodes (if possible). The corresponding routine
in~\texttt{igraph} is~\texttt{barabasi.game(5000, power = 1, m = k/2,
directed = FALSE)}.

\begin{sidewaysfigure}
  \includegraphics[width = \linewidth]{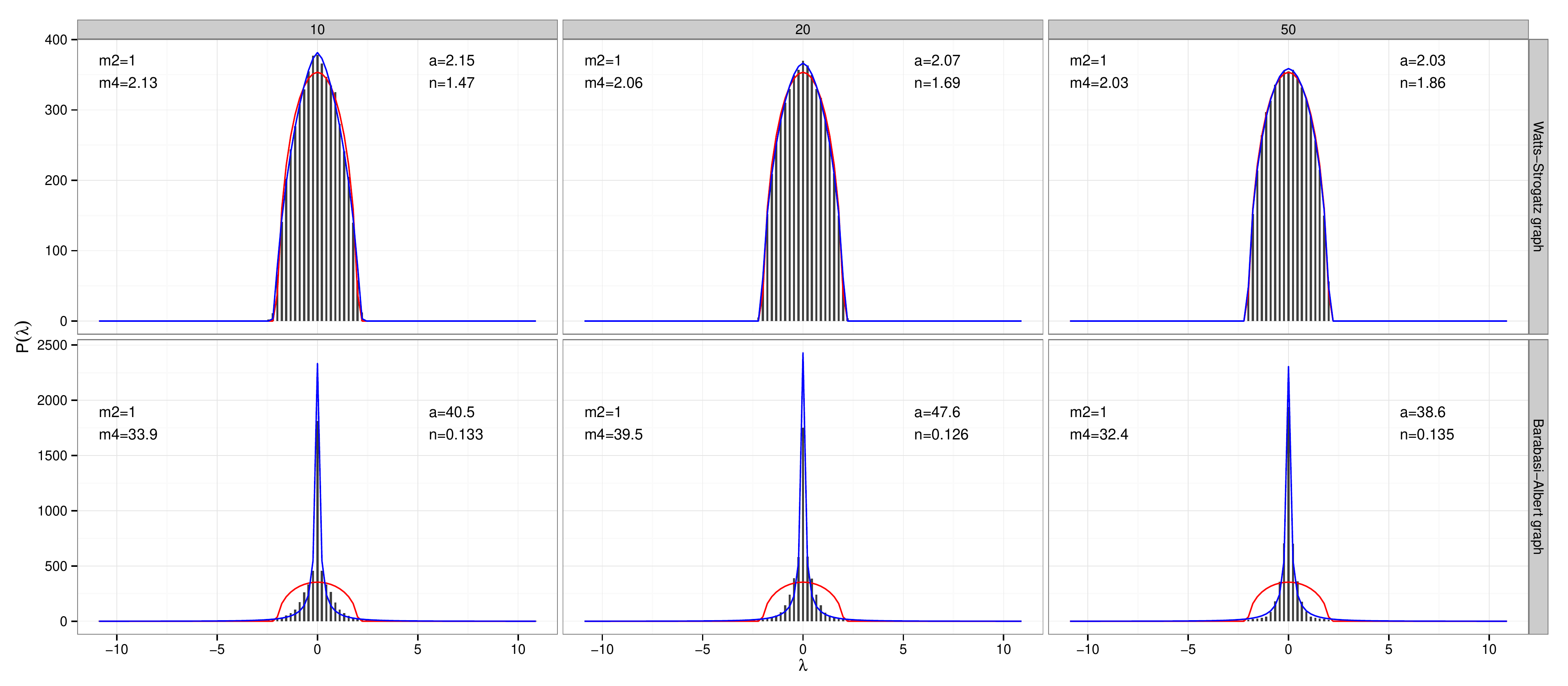}
  \caption{\textit{As main text Figure 1, but using different models
  to construct the networks.}}
  \label{OtherSymm}
\end{sidewaysfigure}

\begin{sidewaysfigure}
 \begin{centering}
  \includegraphics[width = 0.8\linewidth]{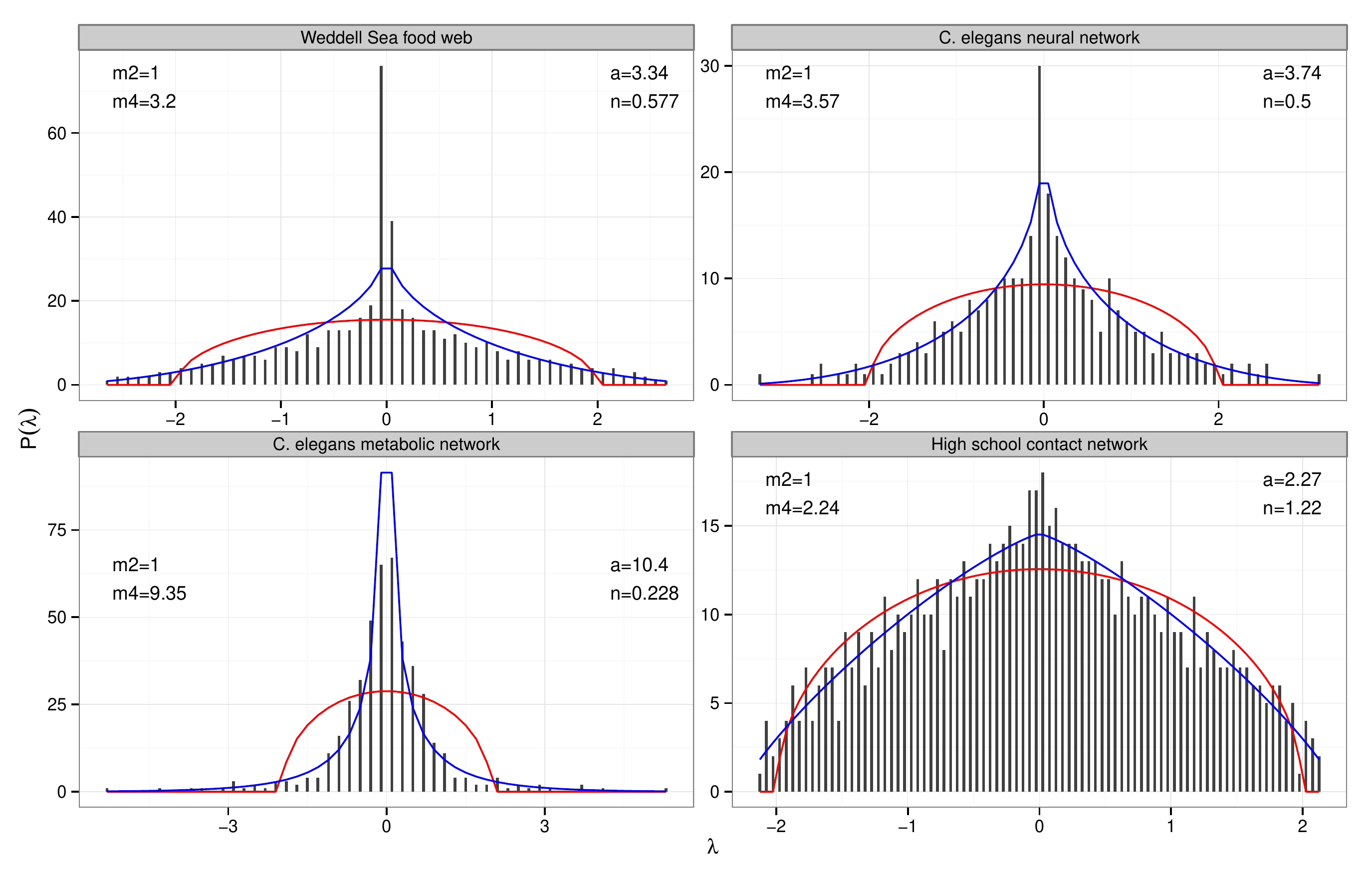}
  \caption{\textit{As main text Figure 1, but using networks of
  biological interest.}}
  \label{BioSymm}
  \end{centering}
\end{sidewaysfigure}

\paragraph*{Symmetric matrices.} We used the empirical networks and
those built using the two models above to construct the symmetric
matrices $M$ (i.e., $M_{ij} = A_{ij}N_{ij}$, where $N$ is symmetric).
The semi-superelliptical distribution captures the eigenvalues of $M$
(Figures~\ref{OtherSymm} and \ref{BioSymm}).

\begin{sidewaysfigure}
  \includegraphics[width = \linewidth]{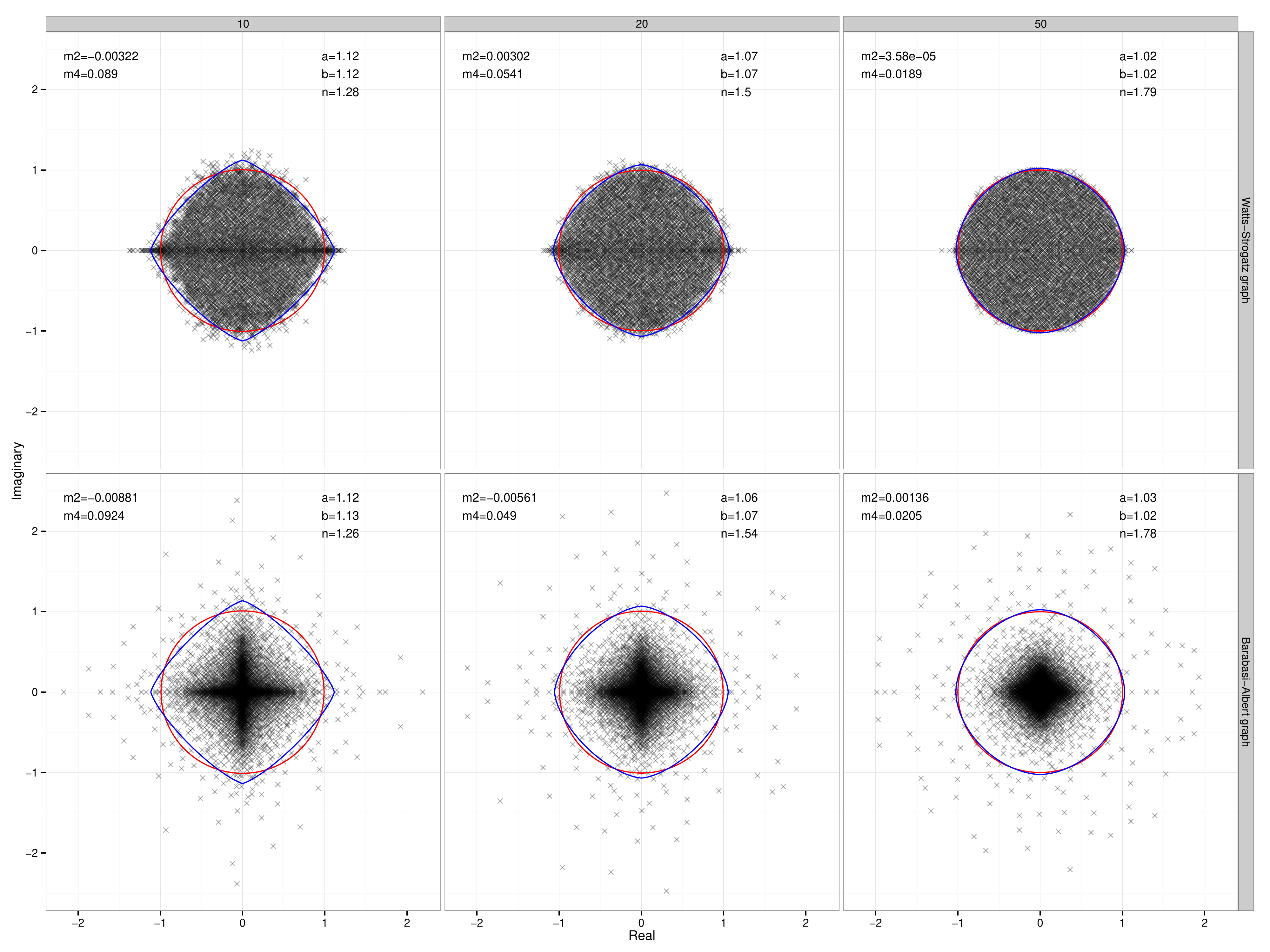}
  \caption{\textit{As main text Figure 2, but using different models
  to construct the networks.}}
  \label{OtherAsymm}
\end{sidewaysfigure}

\begin{sidewaysfigure}
\begin{centering}
  \includegraphics[width = 0.75\linewidth]{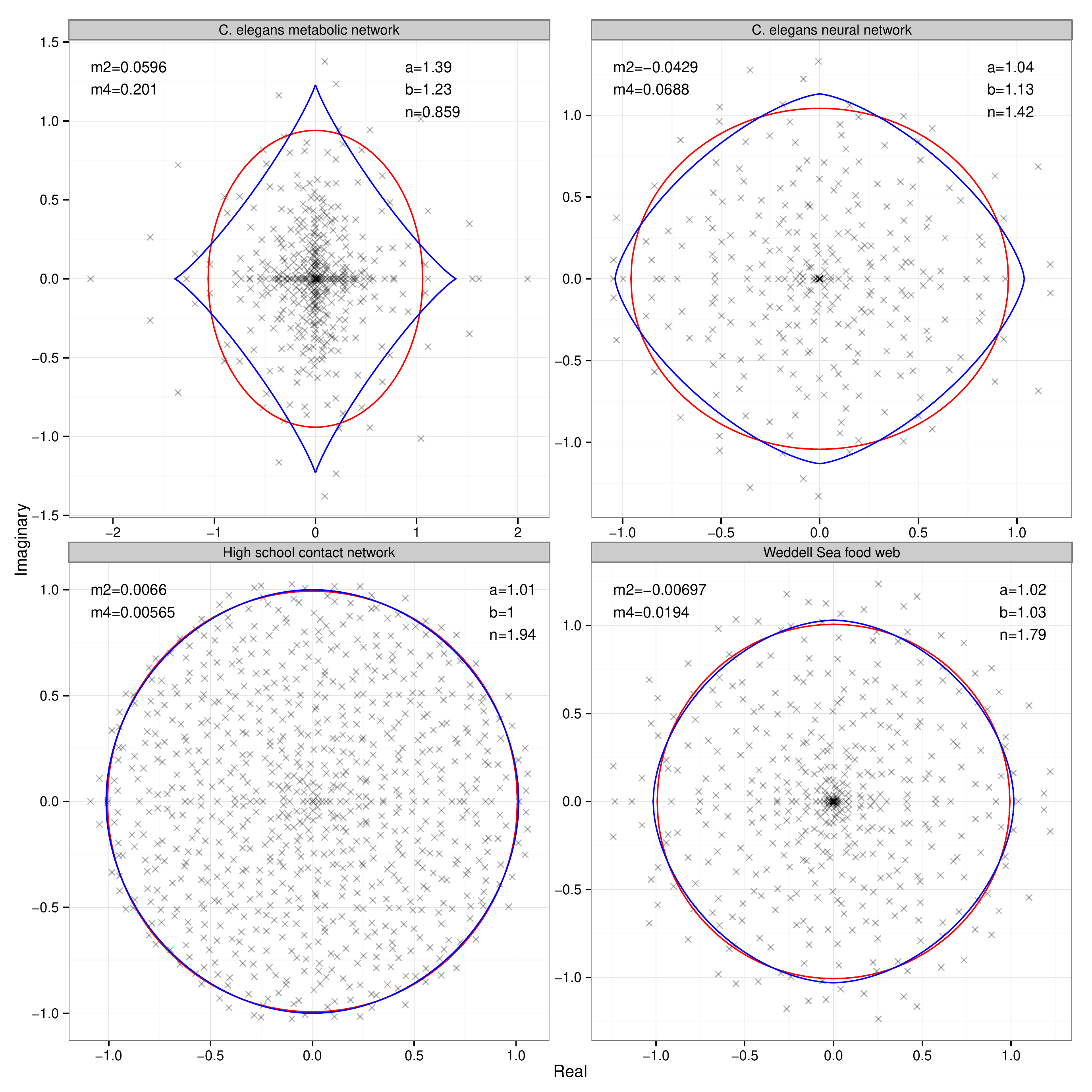}
  \caption{\textit{As main text Figure 2, but using networks of
  biological interest.}}
  \label{BioAsymm}
\end{centering}
\end{sidewaysfigure}

\paragraph*{Asymmetric matrices.} We repeated the exercise using
  asymmetric matrices, and found, that the superelliptical
  distribution describes the eigenvalues quite poorly---as
  expected---, with the exception of the Watts-Strogatz model and the
  High school contact network (Figures~\ref{OtherAsymm}
  and \ref{BioAsymm}).

\begin{figure}
\begin{centering}
  \includegraphics[width = 0.8\linewidth]{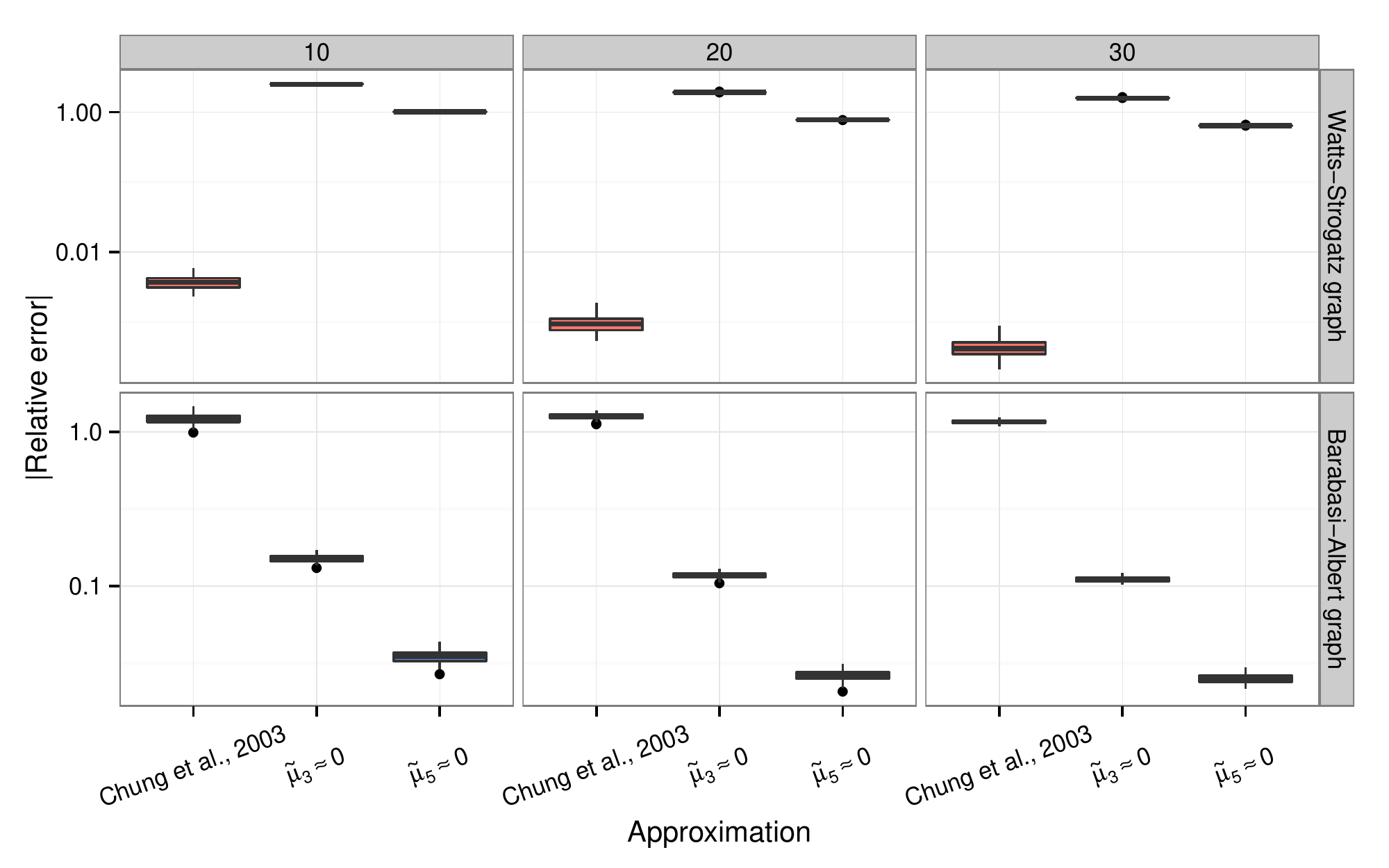}
  \caption{\textit{As main text Figure 4, but using different models.}}
  \label{AdjNRnd}
  \end{centering}
\end{figure}

\begin{figure}
  \includegraphics[width
  = \linewidth]{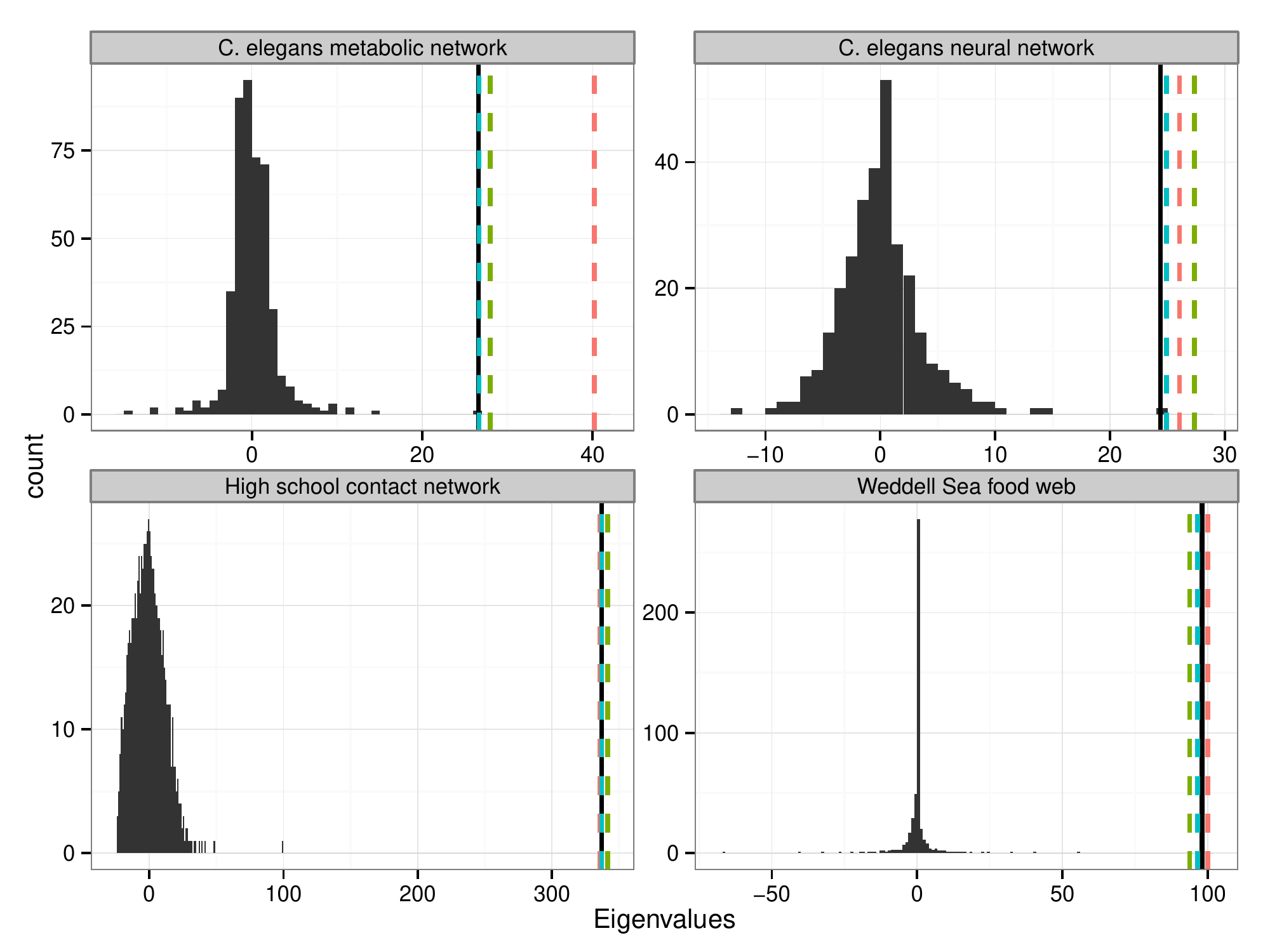} \caption{\textit{The
  spectrum of the adjacency matrix of four networks of biological
  interest. The solid black line marks the value of $\lambda_1$. The
  dashed lines represent the approximations of Chung \textit{et al.}
  (red) and those obtained by setting $\tilde{\mu}_3 \approx 0$
  (green) and $\tilde{\mu}_5 \approx 0$ (turquoise).}}  \label{AdjBio}
\end{figure}

\paragraph*{Approximating $\boldsymbol \lambda_1.$} Finally, we
  constructed a hundred $1000 \times 1000$ adjacency matrices for each
  choice of $k$ using the two models and attempted to approximate
  $\lambda_1$. For the Watts-Strogatz model, the approximation of
  Chung \textit{et al.} is clearly superior, while for the
  Barab\'asi-Albert model and the empirical networks the new
  approximations introduced here produce better results
  (Figure~\ref{AdjNRnd} and \ref{AdjBio}).

\section*{Code}
Code performing the analysis described here is freely available for
download in the public repository:

\noindent \texttt{
https://bitbucket.org/AllesinaLab/superellipses}.

\section*{Acknowledgments}
Research supported by NSF DEB \#1148867. We thank J. Grilli and
   P. Staniczenko for comments.

\clearpage
\bibliography{Biblio}

\end{document}